\documentclass[screen]{acmart}

\usepackage{url}
\usepackage{caption}
\usepackage{subcaption}
\usepackage{array}
\AtBeginDocument{%
  \providecommand\BibTeX{{%
    \normalfont B\kern-0.5em{\scshape i\kern-0.25em b}\kern-0.8em\TeX}}}

\setcopyright{acmcopyright}
\copyrightyear{2022}
\acmYear{2022}
\acmDOI{XXXXXXX.XXXXXXX}

\newcommand{\classname}[1]{\texttt{#1}}

\begin{document}

\title{Simulating the Impact of Dynamic Rerouting on Metropolitan-Scale Traffic Systems}

\author{Cy Chan}
% \authornote{}
\email{cychan@lbl.gov}
\orcid{0000-0001-6881-827X}
\affiliation{%
  \institution{Lawrence Berkeley National Laboratory}
  \streetaddress{1 Cyclotron Rd.}
  \city{Berkeley}
  \state{California}
  \country{USA}
  \postcode{94122}
}

\author{Anu Kuncheria}
\email{anu_kuncheria@berkeley.edu}
\author{Jane Macfarlane}
\email{janemacfarlane@berkeley.edu}
\affiliation{%
  \institution{University of California, Berkeley}
  \streetaddress{109 McLaughlin Hall}
  \city{Berkeley}
  \state{California}
  \country{USA}
  \postcode{94122}
}

% \renewcommand{\shortauthors}{Chan, et al.}

\begin{abstract}
The rapid introduction of mobile navigation aides that use real-time road network information to suggest alternate routes to drivers is making it more difficult for researchers and government transportation agencies to understand and predict the dynamics of congested transportation systems. Computer simulation is a key capability for these organizations to analyze hypothetical scenarios; however, the complexity of transportation systems makes it challenging for them to simulate very large geographical regions, such as multi-city metropolitan areas. In this paper, we describe the Mobiliti traffic simulator, which includes mechanisms to capture congestion delays, timing constraints, and link storage capacity constraints.  The simulator is designed to support distributed memory parallel execution and be scalable on high-performance computing platforms.  We introduce a method to model dynamic rerouting behavior with the addition of vehicle controller actors and reroute request events.  We demonstrate the potential of the simulator by analyzing the impact of varying the population penetration rate of dynamic rerouting on the San Francisco Bay Area road network.  Using high-performance parallel computing, we can simulate a day of the San Francisco Bay Area with 19 million vehicle trips with 50 percent dynamic rerouting penetration over a road network with 0.5 million nodes and 1 million links in less than three minutes.  We present an analysis of system-level impacts when changing the dynamic rerouting penetration rate and examine the varying effects on different functional classes and geographical regions.  Finally, we present a validation of the simulation results compared to real world data.
\end{abstract}

%   <concept>
%       <concept_id>10010147.10010341.10010349.10010354</concept_id>
%       <concept_desc>Computing methodologies~Discrete-event simulation</concept_desc>
%       <concept_significance>500</concept_significance>
%       </concept>
%   <concept>
%       <concept_id>10010147.10010341.10010349.10010362</concept_id>
%       <concept_desc>Computing methodologies~Massively parallel and high-performance simulations</concept_desc>
%       <concept_significance>500</concept_significance>
%       </concept>
%   <concept>
%       <concept_id>10010147.10010341.10010349.10010355</concept_id>
%       <concept_desc>Computing methodologies~Agent / discrete models</concept_desc>
%       <concept_significance>500</concept_significance>
%       </concept>
%  </ccs2012>

% \ccsdesc[500]{Computing methodologies~Discrete-event simulation}
% \ccsdesc[500]{Computing methodologies~Massively parallel and high-performance simulations}
% \ccsdesc[500]{Computing methodologies~Agent / discrete models}

% %%
% %% Keywords. The author(s) should pick words that accurately describe
% %% the work being presented. Separate the keywords with commas.
% \keywords{large-scale transportation simulation, dynamic vehicle rerouting, high-performance computing, parallel discrete event simulation, actor-based modeling}

\maketitle

\section{Introduction}
Active traffic management strategies using changeable message signs or broadcast media are regularly used by traffic management centers to present alternate routes to drivers when abnormal events occur on the roadways. Shifting traffic onto alternate routes maximizes the efficiency and capacity of the network and increases safety by reducing secondary vehicle accidents due to the unexpected congestion. With highways, parallel corridors that handle the rerouted traffic must be available with adequate capacity for rerouting to be successful. This requires traffic centers, with humans in the loop, to assess the situation, have knowledge of the parallel routes via predefined control strategies, and implement an active control plan. In the past five years, smart phone navigation apps have gained popularity and serve a similar role. These applications use their current understanding of congestion to compute new routes in which the travel time is shorter for the user of the device. The congestion information is generated by aggregating the speeds and locations of all of the devices that the app is monitoring. A road congestion estimate is created using road network information and other traffic information, such as historical traffic information, and combined with the current user’s destination. If the projected congestion of the user’s current route significantly impacts their expected travel time, the app may suggest a new route to the user with a shorter travel time. As such, these navigation applications are also serving as active traffic managers.

The introduction of these new active traffic managers makes it more difficult for researchers and government transportation agencies to understand and predict the dynamics of congested transportation systems. Traffic simulation is a key capability for these organizations to analyze different scenarios and predict the potential impacts of different infrastructure changes, policies, or control strategies. However, the complexity of transportation systems makes them extremely difficult to simulate at urban scale. As such, little is known about how to control and actively manage large scale road networks in the presence of significant congestion, particularly with the active management now being implemented by different agents, such as smart phone apps and traffic centers. We do know that providing information about congested states and having a variety of sources provide new travel routes will likely create unpredictable patterns and dynamic variation.
Control strategies implemented through traffic management can include infrastructure modifications such as signal-phase-timing adjustments of signals to redirect traffic and expedite congestion mitigation. These control decisions are determined by estimating the impact of the rerouting action. To-date, these do not include rerouting that occurs as a result of drivers following an app’s routing information. This was made evident when during an evacuation event, drivers were venturing into dangerous situations with routes that were not informed by road closures associated with the event \cite{noauthor_california_nodate}. 

In the last two decades, transportation network simulation has increased in popularity for emulating driving behaviors, predicting traffic dynamics and predicting impacts of control strategies. Applied models can broadly be categorized into both equilibrium models, often referred to as \textbf{traffic assignment} models, and non-equilibrium models, often referred to as \textbf{simulation models} \cite{behrisch_comparing_2010,rojo_evaluation_2020}. The simulation models complement traffic assignment, which is one of the essential tools that planners use for estimating congestion.  With a reliable origin-destination demand matrix, an accurate traffic assignment generates optimized traffic-state predictions \cite{rojo_evaluation_2020}.  While equilibrium models have been very popular in the past due to mathematical clarity, their main drawback is the inability to capture the congestion-dependent evolution of a driver's route \cite{kim_effects_2009}. This limitation is overcome with non-equilibrium models that allow for modeling route guidance dynamics and unexpected events such as incidents or evacuation strategies \cite{noauthor_traffic_analysis_toolbox_nodate, behrisch_comparing_2010}.

\begin{figure}[h]
    \centering
    \includegraphics[width=0.75\textwidth]{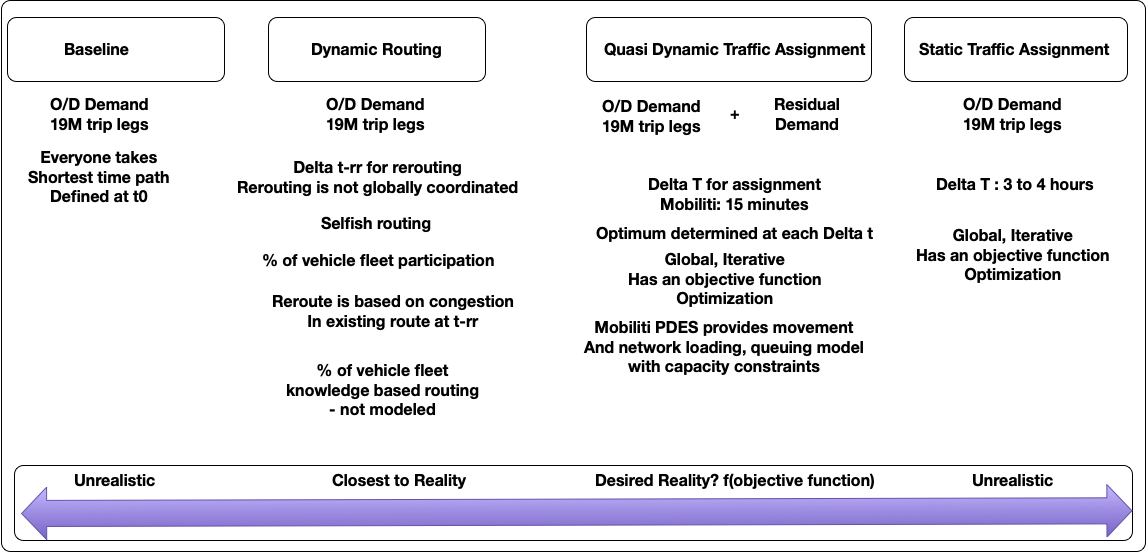}
    \caption{Four traffic models for large scale network modelling that can be employed for a wide array of transportation planning projects. While the first two employ non-equilibrium traffic assignment modelling, the last two use equilibrium based traffic assignment for route choice decisions.}
    \label{fig:model_frameworks}
\end{figure}

In order to predict the emergent traffic dynamics that result from congestion-dependent rerouting and unexpected events, we have extended our parallel discrete event simulation of large-scale traffic dynamics \cite{chan_mobiliti_2018} with new dynamic vehicle rerouting capabilities.  To understand the context of this paper, Figure~\ref{fig:model_frameworks} describes four different traffic models we have developed for large-scale network modeling, each with a different set of algorithms. Details of the baseline model can be found in our previous paper\cite{chan_mobiliti_2018} and the quasi-dynamic traffic assignment model will be described in a forthcoming publication \cite{DBLP:journals/corr/abs-2104-12911}. 
In this paper, we focus on the second model with dynamic routing which we believe best captures realistic emergent traffic dynamics. Note that we acknowledge that there are a percentage of routes that are neither shortest free speed routes nor dynamically routed in the real world. We denote them as \textit{knowledge-based} routes, representing routes that are arbitrarily chosen by a user, such as a route chosen to include a caf\'e stop en route to work. At this time, we do not model these types of routes; thus, all 19M routes are either shortest free speed routes or dynamically routed.

The ultimate goal of active traffic management is to drive the system towards an optimized state, which is the focus of most traffic assignment algorithms. But in reality, reaching equilibrium is highly unlikely regardless of the attempted active management strategies. As such, a mechanism for simulating and analyzing hypothetical scenarios that will reveal emergent dynamics with prescribed dynamic routing algorithms would be extremely valuable for those designing active traffic management systems. With increased urbanization in cities today, understanding how to actively manage the dynamics on the road network is becoming increasingly urgent, not only from a loss in productivity perspective but from a fuel consumption perspective. Our research focus is to breakdown the computational barriers of simulating large scale road network dynamics in order to investigate the impacts of active traffic management strategies and reduce the multifaceted cost of the increasing congestion in our cities.

This paper presents a methodology for representing the transportation system with dynamic rerouting capabilities in a scalable actor-based parallel discrete event simulation (PDES) formalism.  We include a description of the core link actor model and its constituent components that calculate vehicle traversal times, timing constraints, and storage capacity constraints.  We also introduce vehicle controller actors that monitor congestion within the traffic system and handle dynamic reroute requests from simulated vehicles.  We demonstrate that the resulting mesoscopic model achieves scalable computational performance for a large system (19 million vehicles over a network with 1 million links representing the San Francisco Bay Area), simulating a day of traffic with 50 percent dynamic rerouting penetration in three minutes on 512 cores of the NERSC Cori computer at Lawrence Berkeley National Laboratory~\cite{cori}.  We provide a rerouting parameter sensitivity analysis and discussion of the simulated impacts of varying the dynamic rerouting penetration rate on various transportation system metrics, and finally present a validation of the simulation results compared to real-world data sources.

\section{Related Work}

There has been much previous work in the area of transportation system modeling and simulation. Previous simulators can be broadly classified as macroscopic, mesoscopic, and microscopic based on the level of detail of the behavior of individual agents simulated. The level of behavior detail in the model typically also correlates to the size of the area under simulation – larger area simulations tend to use macroscopic models, while simulations focusing on small areas use more detailed microscopic models. DTA solvers and dynamic simulators such as MATSim \cite{noauthor_matsimorg_nodate}, POLARIS \cite{auld2016polaris}, DTALite \cite{zhao_dynamic_2017}, DynaMIT \cite{ben2002real}, DynaSmart \cite{mahmassani_dynamic_1998}, Aimsun~\cite{aimsun}, SUMO \cite{lopez2018microscopic}, INTEGRATION \cite{rakha_integration_2012}, BEAM \cite{sheppard_modeling_2017}, Ugirumurera et. al. \cite{ugirumurera_unified_2018}, MANTA \cite{yedavalli2021microsimulation}, and many others fall along the spectrum of different modeling approaches, target problem scales, modeling fidelities, and output (e.g., traffic assignment versus simulation).  The level of fidelity targeted by our Mobiliti simulator falls somewhere between microscopic and mesoscopic since it does not resolve lane-changing or vehicle following behavior, but it does resolve individual vehicle movements and rerouting decisions.

There is a large body of previous work in the area of parallel discrete event simulation (PDES, see \cite{fujimoto_parallel_1990} for an overview). In particular, seminal work by Chandy and Misra \cite{chandy_distributed_1979}, Bryant \cite{abrams_common_1989}, and Jefferson \cite{Jefferson:1985:VT:3916.3988} laid the groundwork for parallel discrete event simulation, which was shown to run efficiently on supercomputers~\cite{Barnes:2013:WSE:2486092.2486134,BauerJr.:2009:STW:1577959.1577971}.  Previous traffic simulators that utilize PDES include those by Perumalla~\cite{perumalla_systems_2006} and Yoginath~\cite{yoginath_reversible_2009}, who used optimistic PDES for modeling traffic grids, though they evaluated their system on smaller-scale synthetic grid networks and used a different mechanism to mediate congestion among vehicles.  Thulasidasan et. al.~\cite{thulasidasan2009accelerating} also modeled road networks using queue-based parallel discrete event simulation, but they did not include adaptive vehicle \textit{rerouting} behavior in their model, where vehicles can respond to congestion by rerouting \textit{in the middle} of (rather than just at the beginning of) their trip.  Furthermore, our paper includes a description of the design and implementation of a \textit{dynamic vehicle rerouting system} with a parameter sensitivity analysis for vehicles and controllers at large scale.

In the area of dynamic re-routing, research by Liang and Wakahara \cite{liang_real-time_2014} showed how proactive dynamic re-routing could reduce average travel time for congested road networks using the SUMO micro simulator on a medium-sized area of London (3,002 links and 332 nodes with 954 vehicles).  Zhao et. al. \cite{zhao_dynamic_2017} gave a theoretical analysis of the dynamics and stability of equilibrium with travelers that reroute depending on cost difference, and demonstrated their results on some small networks (up to 31 nodes and 40 links with three OD pairs).  Similarly, \cite{wang2021incentive} provides a theoretical treatment of distributed multi-agent route selection problem with incentives, with numerical examples of their approach on the Sioux Falls road network (24 nodes).  In \cite{tseng2021improved}, the authors utilize SUMO and OMNeT++ to model an area of Brooklyn with 380 nodes and 474 links, with a traffic demand of 2,500 vehicles.  Their approach is similar to ours (albeit on a smaller network) in that they utilize a distributed cloud-based vehicle control system, where sensors detect congestion, and routes are suggested to avoid the congestion.  In \cite{kim2020analysis}, the authors utilize a combination of SUMO, Veins, and OMNeT++ tools to analyze the impact of rerouting on a 2x2 grid map (9 nodes, 24 links).  In \cite{qurashi2020modeling}, the authors describe a vanpool scheduler architecture for dynamic rerouting.  They evaluate their approach on a network of the Munich city center, where the focus is more on responding to stochastic vanpool requests rather than dynamic congestion effects.  In \cite{kucharski2019simulation}, they investigate the information comply model to simulate how drivers can react to information about an event and evaluated their approach on a network with approximately 1000 links.  There has also been related work in the area of route selection in the context of dynamic traffic assignment (e.g., \cite{auld2019agent}).  Dynamic traffic assignment approaches typically model converged dynamic driver behavior adapted to daily congestion patterns, rather than more reactive dynamic rerouting scenarios that explore how traffic can respond to unexpected events.  Our approach differs from these previous works in that we leverage high-performance computing techniques to simulate the effect of dynamic rerouting on much larger road networks (on the order of hundreds of thousands to millions of links) with tens of millions of vehicle trips per day to resolve system-wide impacts on large urban regions.

% A related area of recent research includes leveraging neural networks to optimize vehicle routes.  Due to the computational complexity of machine learning approaches, these efforts have been focused on relatively small road networks.  In \cite{lee2020intelligent}, they explore the application of adaptive rerouting to a materials movement inside semiconductor fabrication facilities with approximately 600 sections.  In \cite{chan2021neural}, the authors utilize SUMO for simulation and focus on imputing missing data on relatively small maps: 2 locations in Malaysia with on the order of hundreds of road links.  In \cite{ho2019improved}, they also use SUMO for simulation and explore a pheromone-based approach with a localized k-shortest path algorithm.  They simulate 5,500 vehicles, on a network of approximately 500 links.  In In \cite{joe2020deep}, they explore model route-based markov decision processes with deep reinforcement learning on a small network with 48 locations, two vehicles, and 22 daily orders.  In \cite{bilgram2021online}, the authors explore the effect of using a Uppaal Stratego algorithm using a small SUMO network with 25 intersections and 60 links.  In \cite{mushtaq2021traffic}, the authors explore optimizations to both traffic signal control and vehicle rerouting on a single signalized intersection area (5 nodes, 8 links) with 1000 vehicles.

Traffic simulation systems use a variety of models and algorithms based on the problem that is being addressed. Geographical scale, modeling fidelity, time horizon, and desired output data all lead to a varied landscape of simulation tools. However, to the best of our knowledge, this paper presents the first simulation framework to model the impact of dynamically rerouting vehicles in a large urban region at scale with millions of nodes and links and millions of vehicles in short execution times, as well as present a rerouting parameter sensitivity discussion and detailed model validation analysis.
\section{MOBILITI SIMULATION}

\begin{figure}
    \centering
    \includegraphics[width=0.75\textwidth]{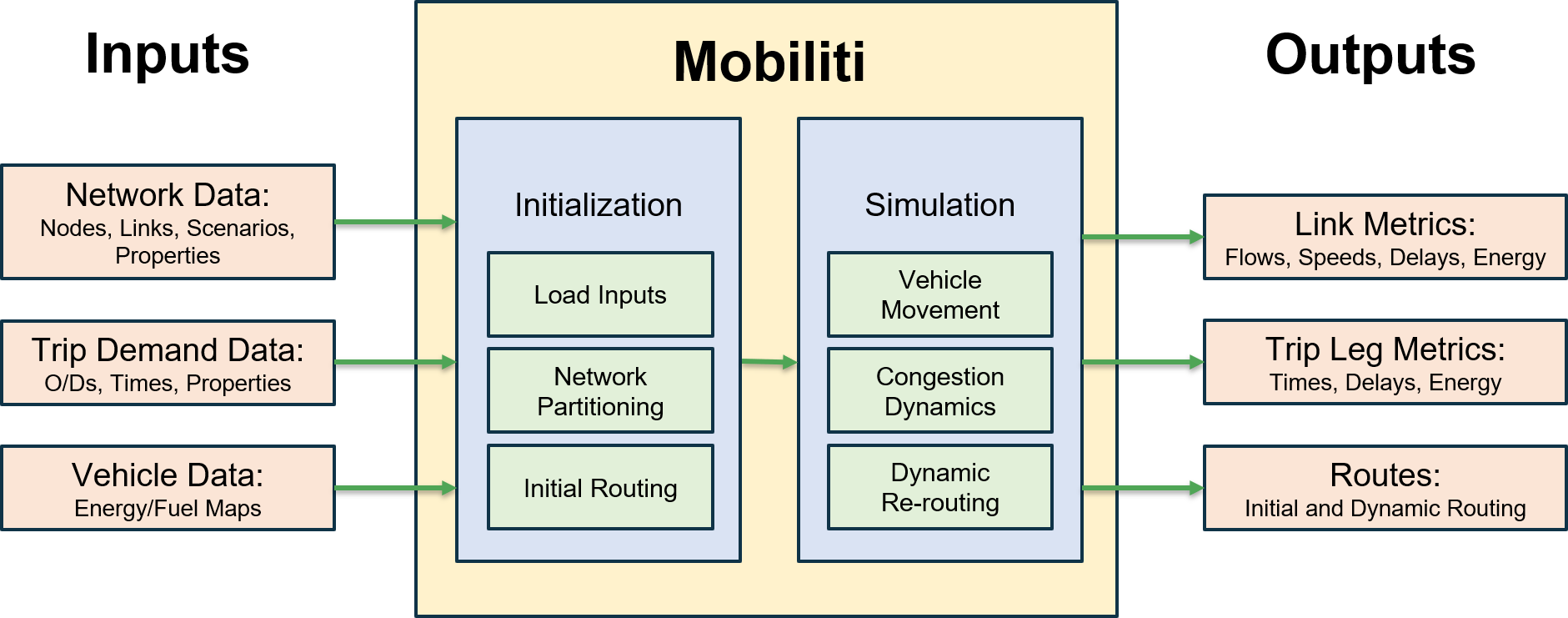}
    \caption{The Mobiliti simulator workflow.  A subset of the inputs and outputs are shown along with a subset of the software stages and modules contained within Mobiliti.}
    \label{fig:workflow}
\end{figure}

\begin{figure}
    \centering
    \includegraphics[width=0.75\textwidth]{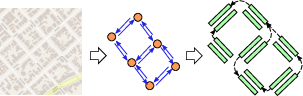}
    \caption{The Mobiliti simulator represents the road network as a collection of link actors that send events to each other representing individual vehicles traversing the network. Vehicles are initialized at their origin nodes and propagated from link to link until they reach their destination. Links are responsible for computing the vehicles' traversal times based on the link's congestion model.}
    \label{fig:model}
\end{figure}

Mobiliti is a distributed-memory, parallel simulation framework that runs on high-performance computing platforms.  It uses an actor-based model and optimistic parallel discrete event simulation (i.e. Time Warp~\cite{Jefferson:1985:VT:3916.3988}) to model the traffic congestion in the network and the rerouting behavior of vehicles in the system. The overall workflow of the simulator is presented in Figure~\ref{fig:workflow}, which shows a concise subset of the inputs, stages of initialization and simulation, and outputs produced by the simulator.  During the simulation phase, the links are modeled as actors and the vehicles are represented as events that are passed from link to link as they travel through the road network. Figure~\ref{fig:model} illustrates the representation of network links as actors passing vehicle events between them.

For background on the simulator’s implementation, please see our previous paper~ \cite{chan_mobiliti_2018}.  This section describes relevant updates to the simulation that improve the accuracy of the link model and enable vehicle dynamic rerouting capabilities.  Specifically, we have added mechanisms inside the link actor to enforce more accurate link timing constraints and storage capacity constraints, and we have added a new set of vehicle controller actors that can be queried by vehicles to compute better routes using the current congested state of the network.

\subsection{Computational Link Model and Representation}
\label{sec:link_model}

\begin{figure}[h]
    \centering
    \includegraphics[width=0.75\textwidth]{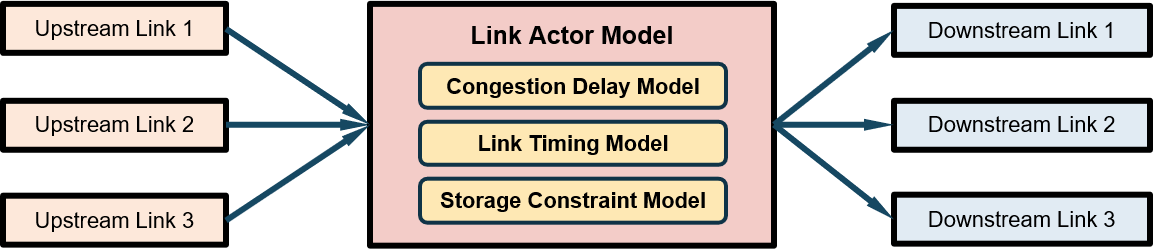}
    \caption{Each link in the simulation is modeled with a Link Actor object that utilizes three sub-models to resolve the various dynamics of the road network.}
    \label{fig:link_actor_model}
\end{figure}

When a vehicle event arrives at a link actor at time $T_0$, the link actor is responsible for determining the simulated time that the vehicle transitions to the next link on its route.  As shown in Figure~\ref{fig:link_actor_model}, the link actor utilizes three sub-models to determine the vehicle's transition time: a flow-based congestion delay model, a link timing model, and a storage constraint model that limits the occupancy on each link based on physical dimensions.

\begin{figure}[h]
    \centering
    \includegraphics[width=0.75\textwidth]{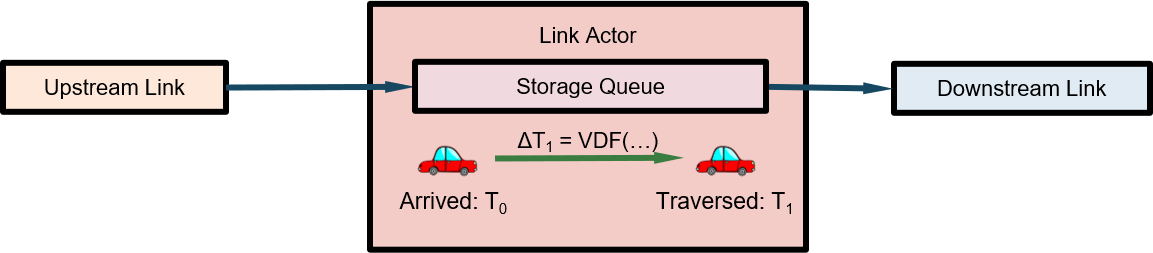}
    \caption{The Congestion Delay Model consists of a vehicle delay function (VDF) that uses link characteristics (such as designated flow capacity) and current activity (such as flow rate) to compute estimated traversal times $\Delta T_1$.}
    \label{fig:delay_model}
\end{figure}

The flow-based congestion delay model (Figure~\ref{fig:delay_model}) uses characteristics of the link (such as the designated flow capacity) and the current activity on the link to estimate the time $\Delta T_1$ it takes for vehicles to traverse from the beginning to the end of the link.  The resulting time $T_1 = T_0 + \Delta T_1$ represents when the vehicle reaches the end of the link; however, the vehicle may be delayed further before actually transitioning to the next link in its route due to additional timing and storage capacity constraints.

\begin{figure}[h]
    \centering
    \includegraphics[width=0.75\textwidth]{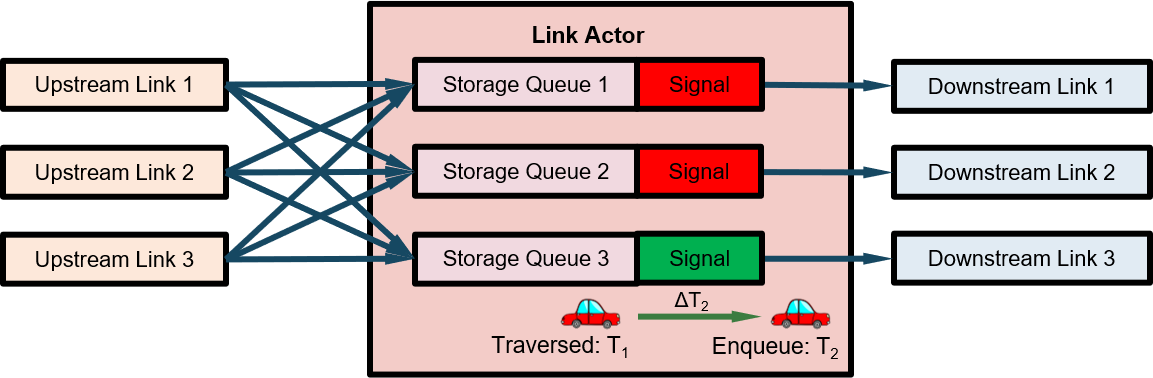}
    \caption{The Link Timing Model assigns transition request times based on minimum vehicle spacing requirements and signal phase timing information.}
    \label{fig:link_actor_queues}
\end{figure}

The link timing model computes the times at which vehicles may legally traverse from the current link to the next.  Separate internal queues are maintained for each maneuver through the intersection (i.e. each downstream link sequence) so that vehicles utilizing different signal phases can be handled independently.  Figure~\ref{fig:link_actor_queues} shows a diagram of the internal queue data structures each link actor utilizes to track the vehicles currently occupying it.  The timing model for each link is fully parameterized so that different, coordinated signal timing plans can be used at multiple intersections in the simulation.

The link timing model actually ensures two things: 1) that vehicles do not transition from link to link at a rate that is faster than physically allowable (determined by the product of vehicle speed, density, and number of lanes), and 2) that vehicles obey the traffic signals (if there is one at this link).  For each queue within the link actor, it keeps track of the previous vehicle's transition time so that the next vehicle traversal maintains a minimum time spacing between transitions to the downstream link.  Furthermore, if the intersection has a traffic signal, vehicles may only transition to the downstream link when the corresponding signal phase is green.  When combined with the minimum spacing requirement, each green phase is modeled as a time interval with a fixed number of consecutive vehicle \textit{slots} during which no more than that fixed number of vehicles may transition to the downstream link.  For each vehicle, the timing model assigns a time $T_2 = T_1 + \Delta T_2$ that obeys the above constraints.  For our current experiments, we utilized the timing model only to enforce the maximum rate for vehicle arrivals at each link since we do not currently have comprehensive signal timing and phase information for the San Francisco Bay Area network.  However, the link capacity parameters used in the vehicle delay functions in the simulator should reflect the lower capacity constraints for links that contain signals.

\begin{figure}[h]
    \centering
    \includegraphics[width=0.75\textwidth]{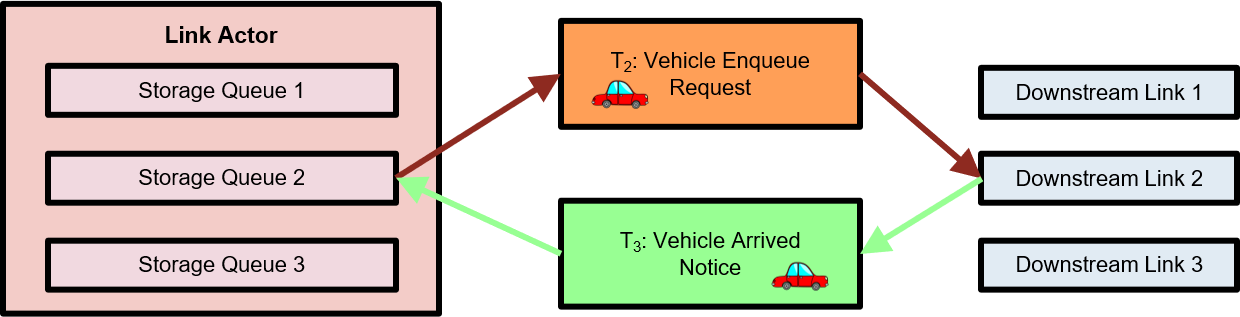}
    \caption{The Storage Capacity Constraint Model ensures that vehicles only transition from the upstream link to a downstream link if the downstream link has sufficient physical storage capacity to accept the vehicle.}
    \label{fig:storage_events}
\end{figure}

Finally, the storage capacity constraint ensures that links do not allow more vehicles to occupy the link than there is physical space.  The storage capacity constraint is mediated by a system of \classname{EnqueueRequest} and \classname{ArrivedNotice} events between the upstream and downstream links.  After a vehicle's preliminary link traversal time $T_2$ is calculated using the congestion delay and timing models, the link actor sends a vehicle enqueue request event to the corresponding downstream link actor at its requested transition time $T_2$.  However, the vehicle remains in the upstream link's storage queues until a response is received.  Only when there is sufficient capacity on the downstream link to accept the incoming vehicle does the downstream link send an \classname{ArrivedNotice} event to the upstream link (at time $T_3 = T_2 + \Delta T_3$) to notify the upstream link that the vehicle has made the transition.  At that time, the upstream link may remove the vehicle from its storage queues, potentially freeing up space to send another \classname{ArrivedNotice} event further upstream.  Figure~\ref{fig:storage_events} illustrates the described protocol that coordinates the transition of vehicles between connected link actors.  The final traversal time for the vehicle over the link is $\Delta T = \Delta T_1 + \Delta T_2 + \Delta T_3$.

\subsection{Dynamic rerouting design}

This section details the simulation mechanism that controls how vehicles dynamically reroute in response to system congestion.  The subset of vehicles that have dynamic rerouting enabled (e.g. those with navigation devices) is specified at program initialization, determining the population rerouting penetration rate.  In order to simulate dynamic rerouting behavior in the system, we implemented an additional set of actors and events that are responsible for coordinating when and how vehicles reroute. These include \classname{VehicleController} actors, \classname{LinkStatusUpdate} events, and vehicle \classname{RerouteCheck} events. These new actors and events are described in the following sections, and relevant parameters are summarized in Table~\ref{tab:reroute_params}.

\begin{table}
\small
\centering
\caption{Tunable dynamic rerouting parameters}
\label{tab:reroute_params}
\begin{tabular}{ccc}
\hline
\noalign{\smallskip}
Parameter & Description & Nominal Value \\
\noalign{\smallskip}
\hline
\noalign{\smallskip}
$t_\text{lsu}$ & Absolute link status update threshold & 60 s \\
$r_\text{lsu}$ & Relative link status update threshold & 1.0 \\
$t_\text{check}$ & Reroute check interval & 300 s \\
$t_\text{delay}$ & Absolute reroute threshold & 120 s \\
$r_\text{delay}$ & Relative reroute threshold & 0.2 \\
\noalign{\smallskip}
\hline
\end{tabular}
\end{table}

\subsubsection{\classname{VehicleController} actors}

\begin{figure}
    \centering
    \includegraphics[width=0.75\textwidth]{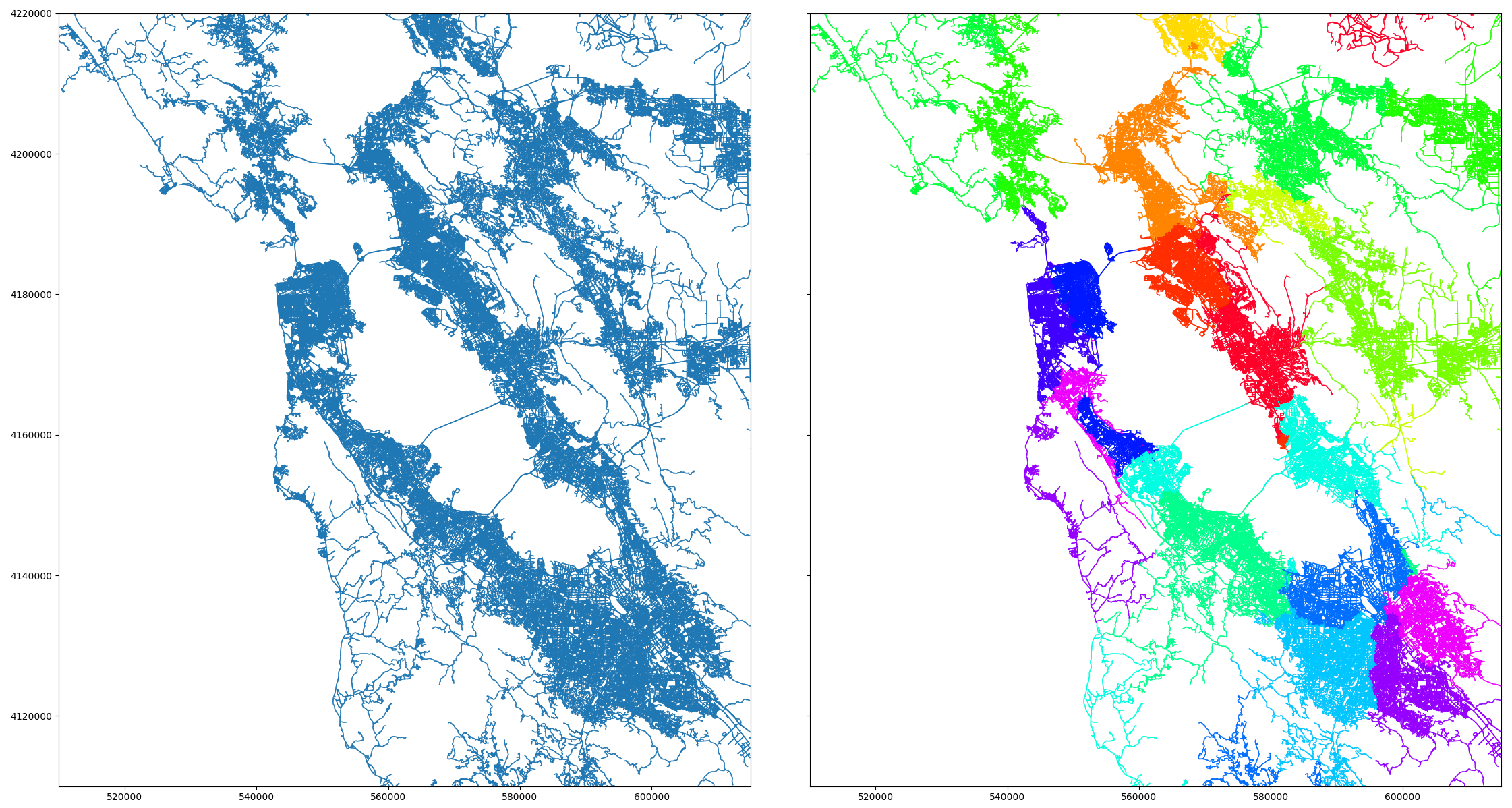}
    \caption{An example partitioning of the SF Bay Area road network into sixteen subgraphs.  Mobiliti partitions the road network into subgraphs and assign one to each computational rank in the simulator. The partitions are load balanced to enable parallel scalability, and each partition is assigned its own \classname{VehicleController} actor to service rerouting requests from vehicles currently in the partition.}
    \label{fig:partitions}
\end{figure}

\classname{VehicleController}s are a new class of actors that can be queried by vehicles to check whether the vehicle should change its route based on current congestion conditions to get to its destination more quickly.
There are multiple controllers instantiated throughout the road network, and since we already partition the road network across simulator ranks (threads) for parallel execution, we use the same partitioning for the \classname{VehicleController}s in our experiments.  
While each individual controller is only responsible for servicing requests originating from vehicles \textit{within} its partition, it maintains current congestion data across the \textit{entire} network, so newly computed routes take the state of the whole system into account.
Figure~\ref{fig:partitions} shows an example of how the San Francisco Bay Area road network could be partitioned into sub-networks.
A distinct \classname{VehicleController} actor is assigned to each sub-network shown in different colors.
When vehicles check whether to reroute, they contact the \classname{VehicleController} assigned to the vehicle’s current road partition.
This method preserves geospatial locality within the model and also has good computational behavior since the \classname{VehicleController}s responsible for routing calculations are parallelized across compute resources in a similar manner that link actors are parallelized.

\subsubsection{\classname{LinkStatusUpdate} events}

\begin{figure}
    \centering
    \begin{subfigure}[h]{0.32\textwidth}
        \includegraphics[width=\textwidth]{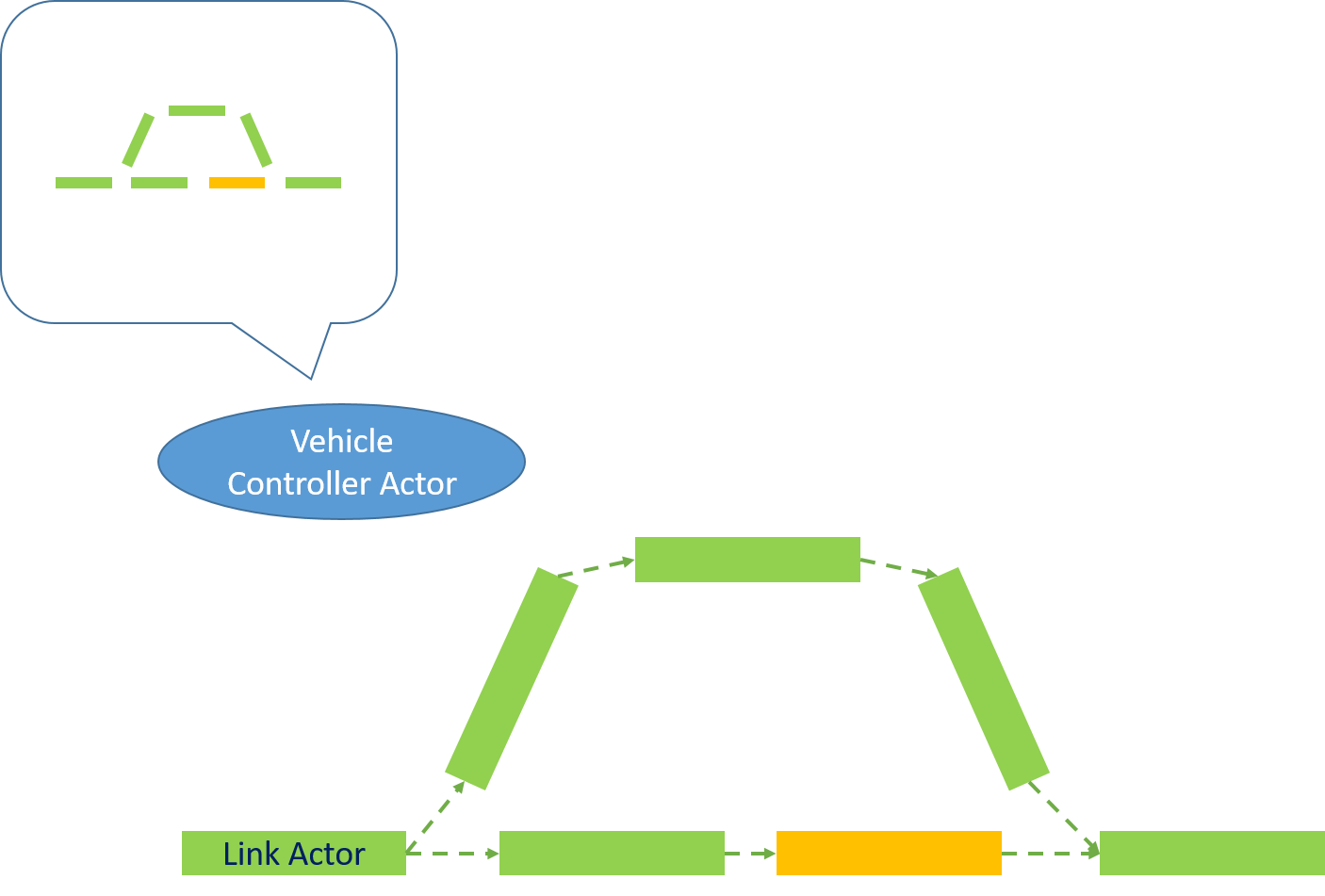}
        \label{subfig:model_0}
        \caption{Initial controller and link states}
    \end{subfigure}
    \hfill
    \begin{subfigure}[h]{0.32\textwidth}
        \includegraphics[width=\textwidth]{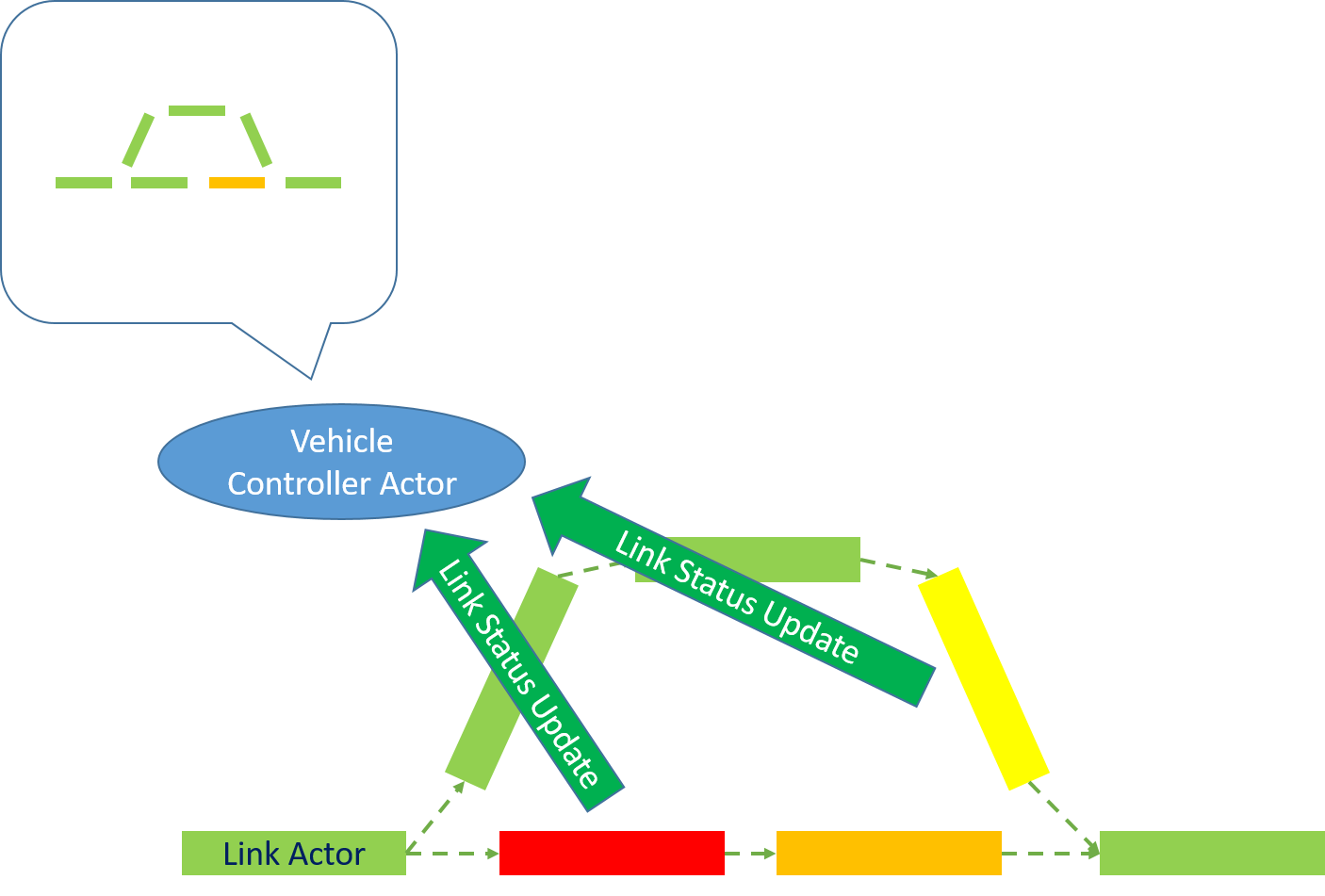}
        \label{subfig:model_1}
        \caption{\classname{LinkStatusUpdate} events}
    \end{subfigure}
    \hfill
    \begin{subfigure}[h]{0.32\textwidth}
        \includegraphics[width=\textwidth]{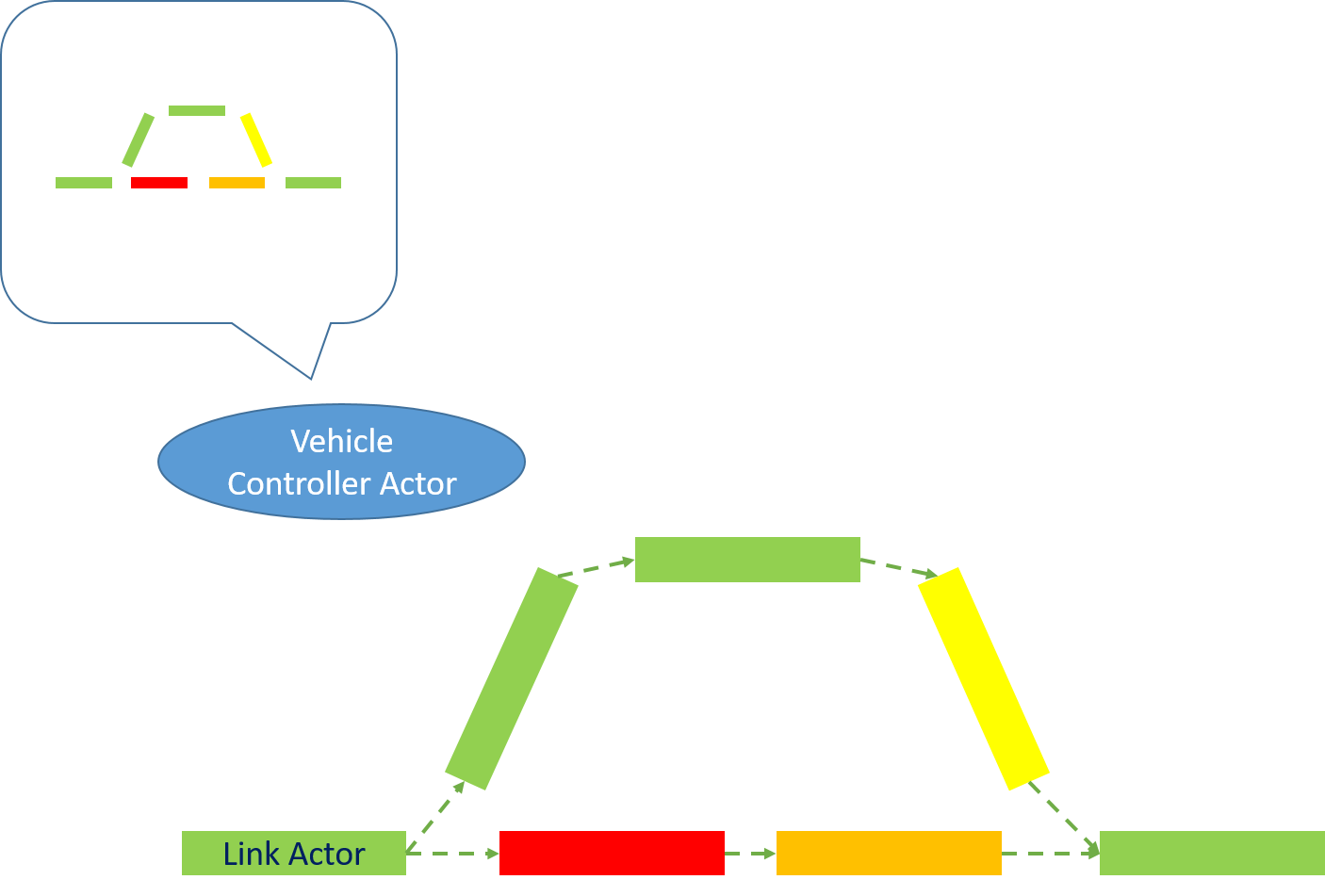}
        \label{subfig:model_2}
        \caption{Updated controller state}
    \end{subfigure}
    \caption{In (a), the \classname{VehicleController} actor’s knowledge of the congestion on the links is up to date. When the congestion on a link changes during simulation, the link actor sends \classname{LinkStatusUpdate} events to the \classname{VehicleController} actors to notify them of its current condition. Figure (b) illustrates an example where two links experience increased congestion and send updates to the \classname{VehicleController} actor. The controller then updates its knowledge of the graph (c) and uses this updated information to calculate routes for rerouting vehicles.}
    \label{fig:link_status_update}
\end{figure}

In order for the \classname{VehicleController}s to have an up-to-date picture of current congestion conditions to make rerouting decisions, each link has a new sub-component that sends updates about its current congestion status to the \classname{VehicleController}s. Figure~\ref{fig:link_status_update} illustrates an example where two link actors send status updates to a \classname{VehicleController}, which then updates its knowledge of the road network’s congested link traversal times.

Every time a vehicle departs a link $l$, the link actor broadcasts an update to \textbf{all} \classname{VehicleController}s with its new congested traversal time if it differs from the previously sent traversal time by at least $\min(t_\text{lsu},r_\text{lsu} \cdot t_f(l))$, where $t_\text{lsu}$ is an absolute threshold, $r_\text{lsu}$ is a relative threshold ratio and $t_f(l)$ is the link's freespeed traversal time.  By allowing some deviation, the thresholds prevent links from sending excessive status updates while ensuring that the values used by the \classname{VehicleController}s for rerouting are still close to the current value experienced on the link.  By making the thresholds tighter, we can increase the frequency that \classname{LinkStatusUpdate} events are sent to the \classname{VehicleController}s, potentially improving rerouting accuracy at the cost of processing more update events.  The sensitivity to these threshold parameters is explored in Section~\ref{sec:parameters}.  Finally, each link actor \textbf{that has active traffic} periodically sends a \textit{heartbeat} update to the vehicle controllers to indicate that it is still servicing traffic at a given speed.  When a vehicle controller has not heard from a link in more than the heartbeat period, it knows the link has not had any traffic, and thus it can purge stale congestion data about the link from its database.

\captionsetup[sub]{font=footnotesize,labelfont={bf,sf}}

\subsubsection{\classname{RerouteCheck} events}

When a vehicle arrives at a link, a vehicle can query the local \classname{VehicleController} with a message that contains the vehicle’s current path $p$ (sequence of links) from its present location to its destination.  The controller estimates the total congested time $t_c(p)$ on the vehicle’s path to its destination using its knowledge of current network congestion. 
% When estimating the contribution of each link to the vehicle’s path delay, the controller may optionally decay each link’s congested value to some historical value since the congestion on a far away downstream link may change by the time the vehicle actually gets to the link.
If the delay on its path $d(p) = t_c(p) - t_f(p)$ exceeds a threshold $\max(t_\text{delay}, r_\text{delay} \cdot t_f(p))$, where $t_\text{delay}$ is an absolute threshold, $r_\text{delay}$ is a relative threshold ratio, and $t_f(p)$ is the freespeed traversal time on path $p$, then the controller calculates a new shortest path $p'$ from the vehicle’s current location to its destination. If the new path time $t_c(p')$ improves the vehicle’s expected arrival time by more than $\max(t_\text{delay}, r_\text{delay} \cdot t_c(p)$, then the vehicle accepts the new path $p'$; otherwise, it continues on its existing path $p$.
We assume a 100 percent compliance rate with the route suggestion given by the \classname{VehicleController} (with no delay) since our simulator does not include human behavior models at this time.  A lower compliance rate may be approximated by adjusting the population rerouting penetration rate accordingly.  
Finally, in order to prevent excessive reroute queries, the vehicle only sends a \classname{RerouteCheck} event if it has been more than $t_\text{check}$ seconds since its last check.
For our experiments, the \classname{VehicleController} actors are homogeneous, simulating a unified network control agency. They could also be configured heterogeneously to simulate the impact of having multiple companies with different (imperfect) network information servicing vehicle reroute requests.  This is a subject for future investigation.

\begin{figure}
    \centering
    \begin{subfigure}[h]{0.32\textwidth}
        \includegraphics[width=\textwidth]{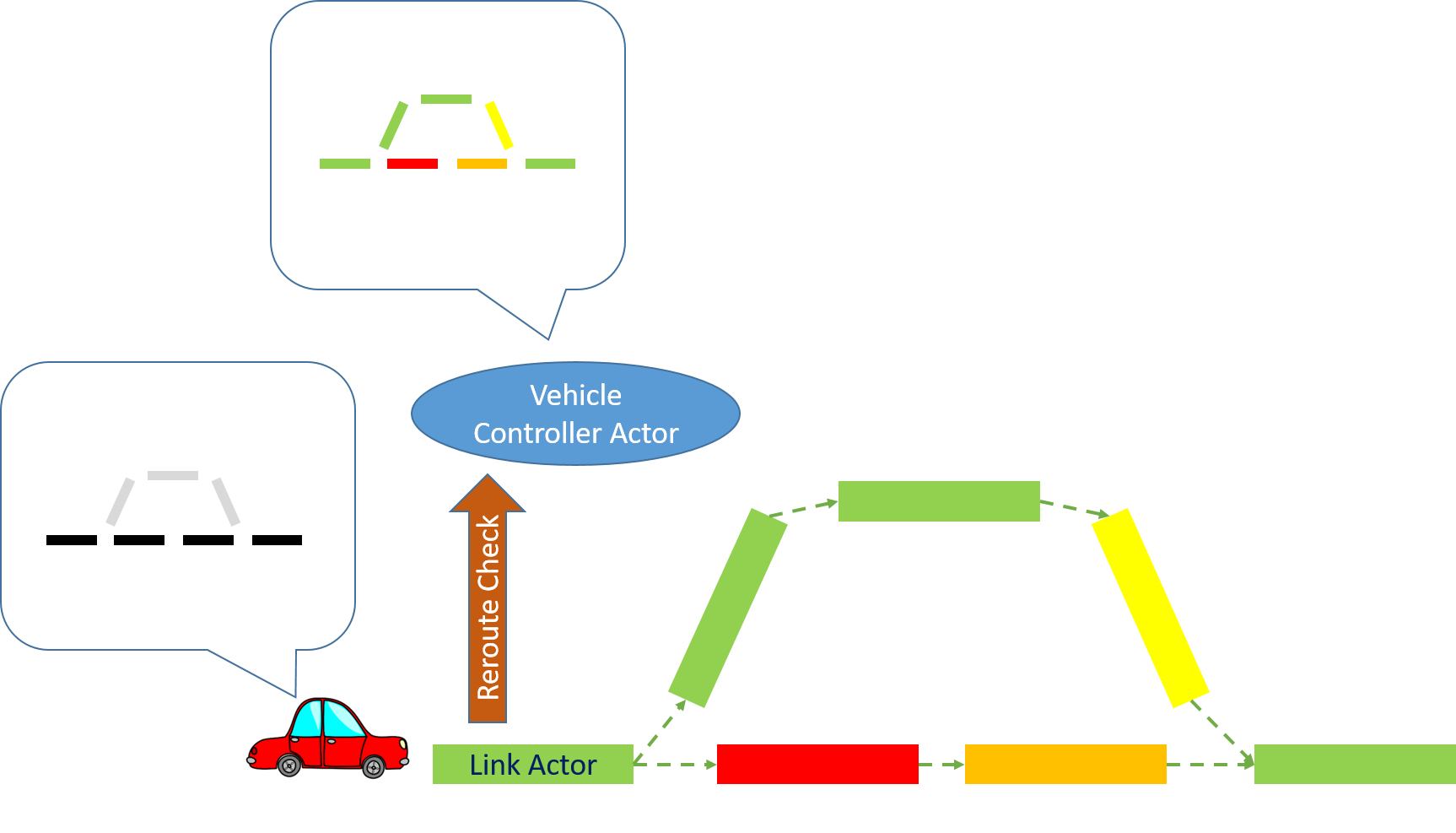}
        \caption{Vehicle arrival and \classname{RerouteCheck} query}
    \end{subfigure}
    \hspace{.1\textwidth}
    \begin{subfigure}[h]{0.32\textwidth}
        \includegraphics[width=\textwidth]{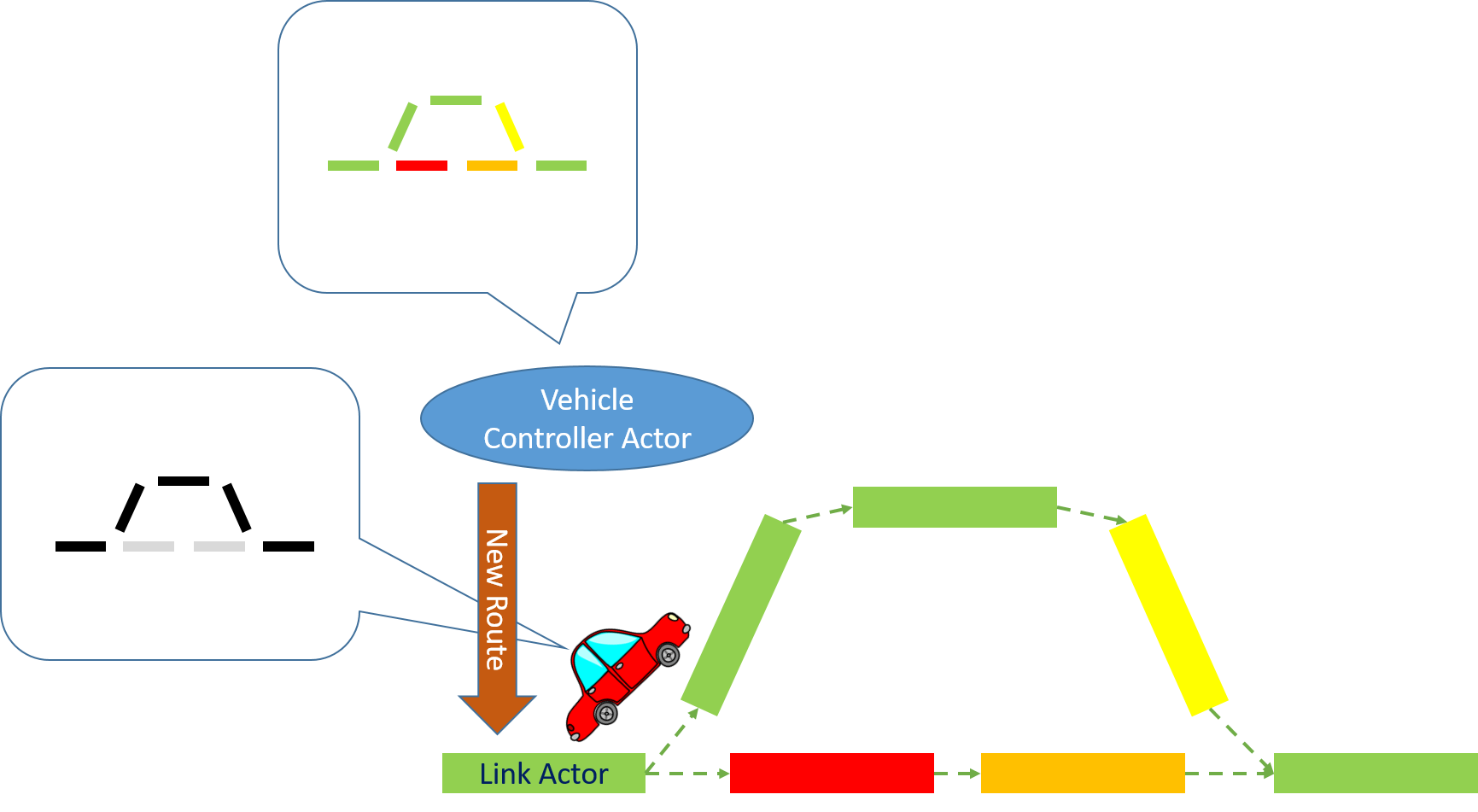}
        \caption{\classname{RerouteCheck} response with new route}
    \end{subfigure}
    \caption{Example \classname{RerouteCheck} event for dynamic rerouting a vehicle based on current congestion conditions. Since the vehicle’s original route is congested, the controller sends an alternate route around the congestion, which the vehicle then takes.}
    \label{fig:sample_subfigures}
\end{figure}

\subsection{Model Parameter Selection and Sensitivity}
\label{sec:parameters}

The various parameters described above and summarized in Table~\ref{tab:reroute_params} control various aspects of the system’s rerouting behavior and can be tuned to explore how the system would hypothetically behave with different parameterizations.  The parameters can be grouped into three categories: 1) $t_\text{lsu}$ and $r_\text{lsu}$ control the frequency of link status updates, where lower values result in more update messages and provide the controllers with more accurate information; 2) $t_\text{check}$ controls the frequency that vehicles contact the controllers to see if they should reroute, where lower values result in more requests; and 3) $t_\text{delay}$ and $r_\text{delay}$ control how aggressive the controllers are at rerouting vehicles, where lower values result in additional reroutes in situations with smaller time saving benefits.

\begin{figure*}
    \centering
    \begin{subfigure}[b]{0.32\textwidth}
        \includegraphics[width=\textwidth]{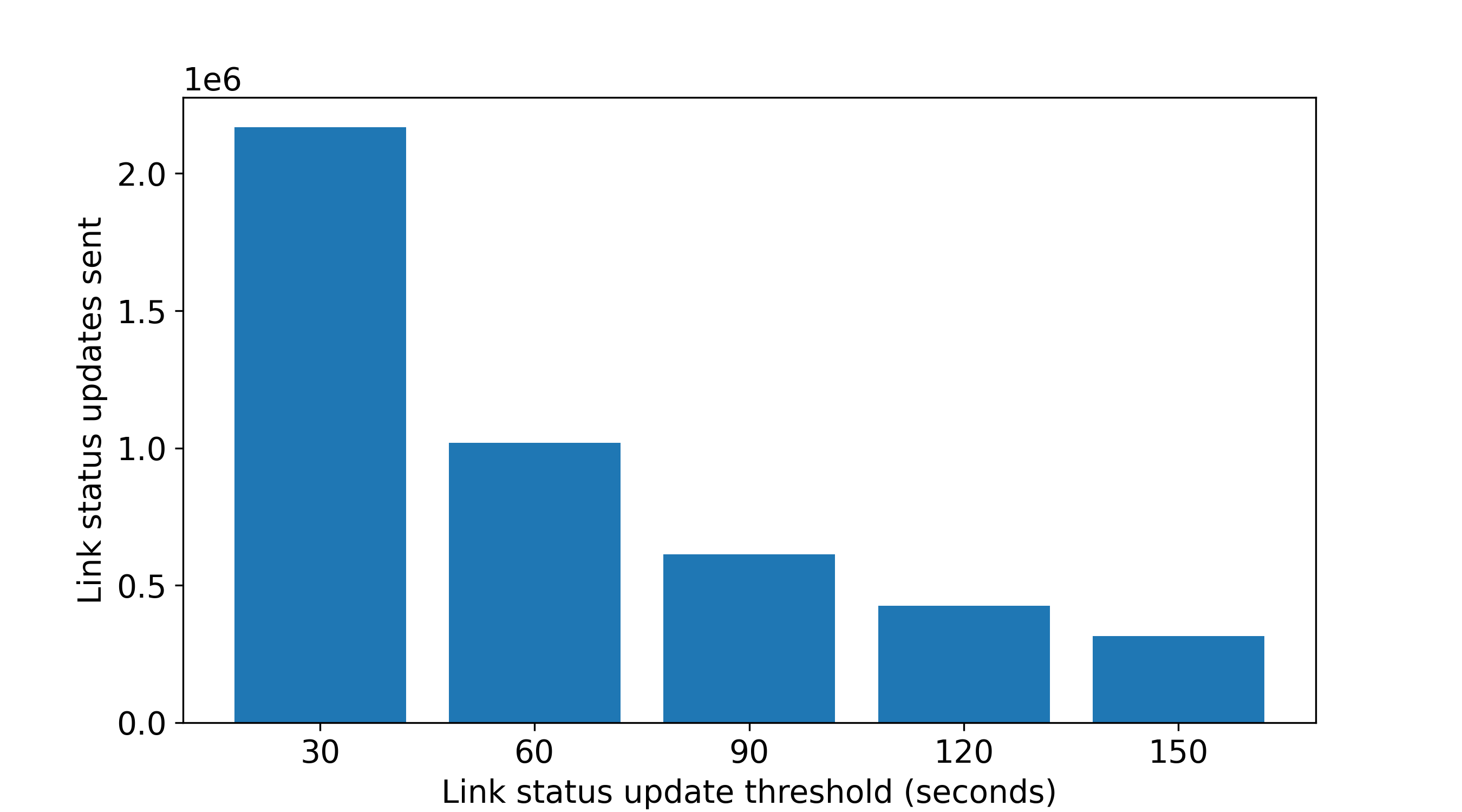}
        \caption{Link Status Updates Sent}
    \end{subfigure}
    \begin{subfigure}[b]{0.32\textwidth}
        \includegraphics[width=\textwidth]{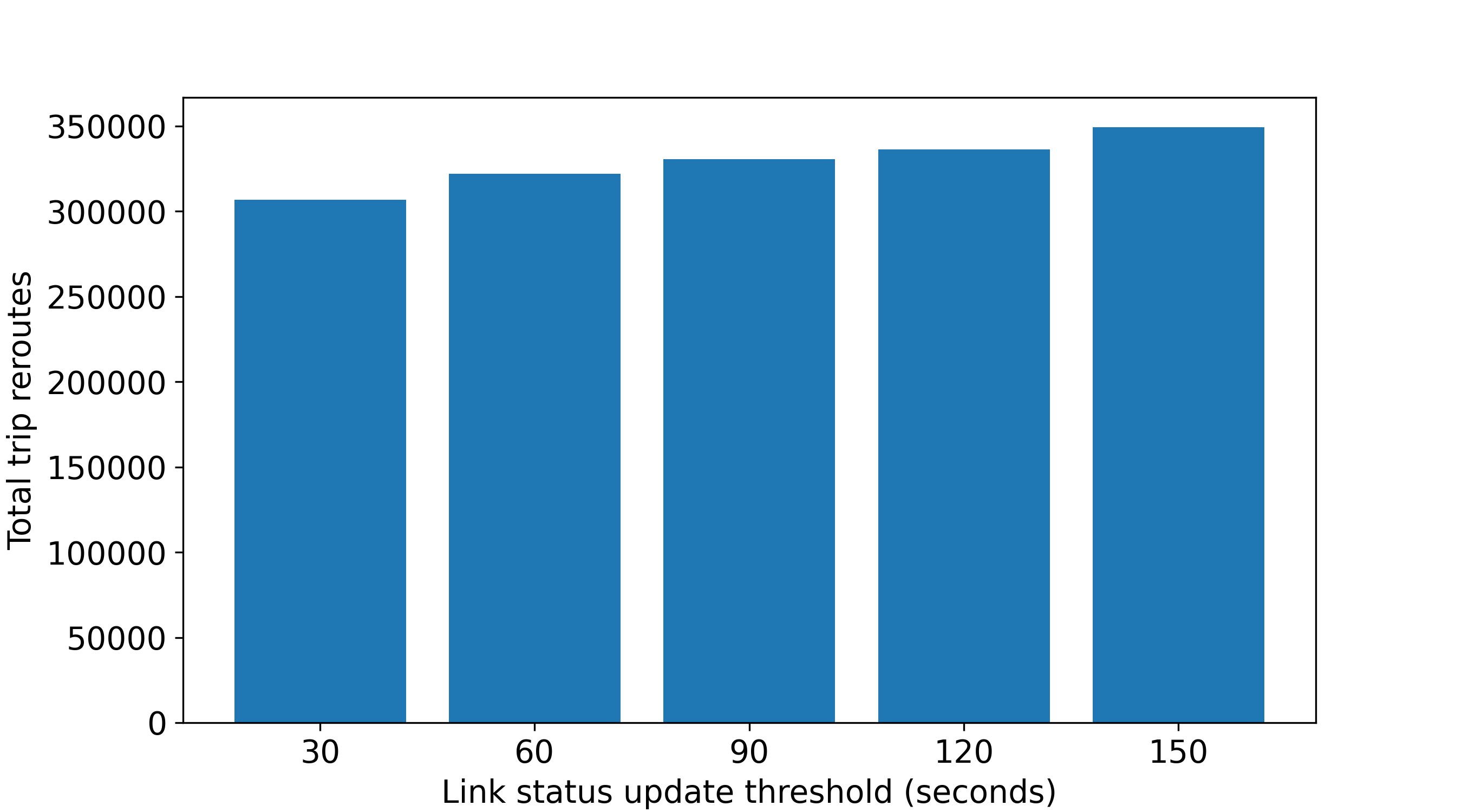}
        \caption{Trip Leg Reroutes}
    \end{subfigure}
    \begin{subfigure}[b]{0.32\textwidth}
        \includegraphics[width=\textwidth]{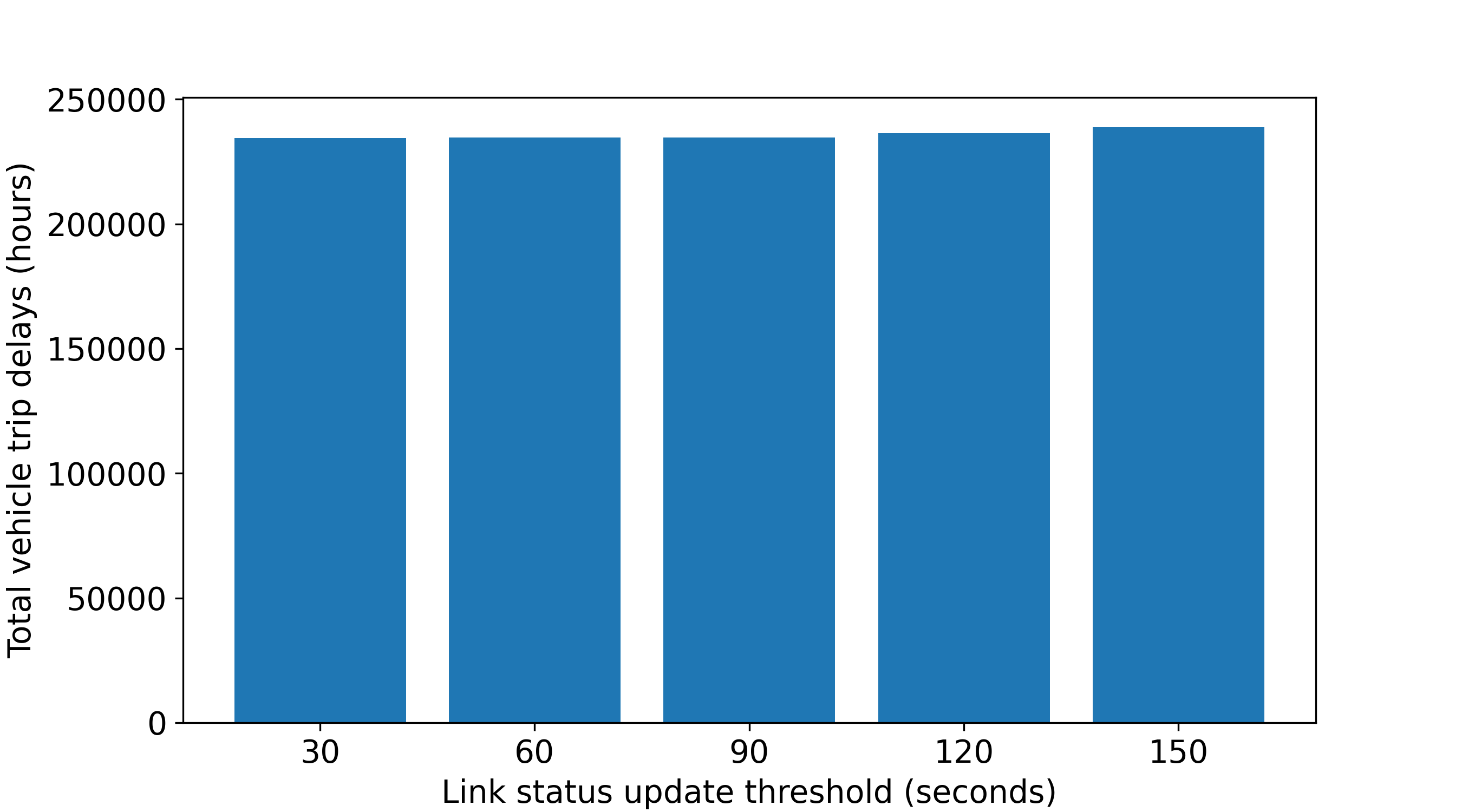}
        \caption{Total (System) Vehicle Delay}
    \end{subfigure}
    \caption{Parameter sensitivity sweep for link status update thresholds. X-axis is the absolute link status update threshold $t_\text{lsu}$, where $r_\text{lsu} = t_\text{lsu} / 60$.}
    \label{fig:lsu_sweep}
\end{figure*}

In order to understand the sensitivity to these parameters, we conducted parameter sweeps for each of these three groups, where we simultaneously varied the values of the parameters in each group, while keeping the other parameters constant to isolate the impact of each group of parameters.  Table~\ref{tab:reroute_params} shows the baseline parameter values.  In the first parameter sweep shown in Figure~\ref{fig:lsu_sweep}, we varied the link status update thresholds: $t_\text{lsu}$ (on the x-axis) and set $r_\text{lsu} = t_\text{lsu} / 60$.  As expected, the number of link status updates sent (a) decreases significantly as the thresholds increase; however, the impact on the number of trip leg reroutes (b) and on the total system delay (c) is relatively small, indicating that there is little benefit from sending very frequent updates, and that the vehicle controllers do a satisfactory job even with coarse information.

\begin{figure*}
    \centering
    \begin{subfigure}[b]{0.32\textwidth}
        \includegraphics[width=\textwidth]{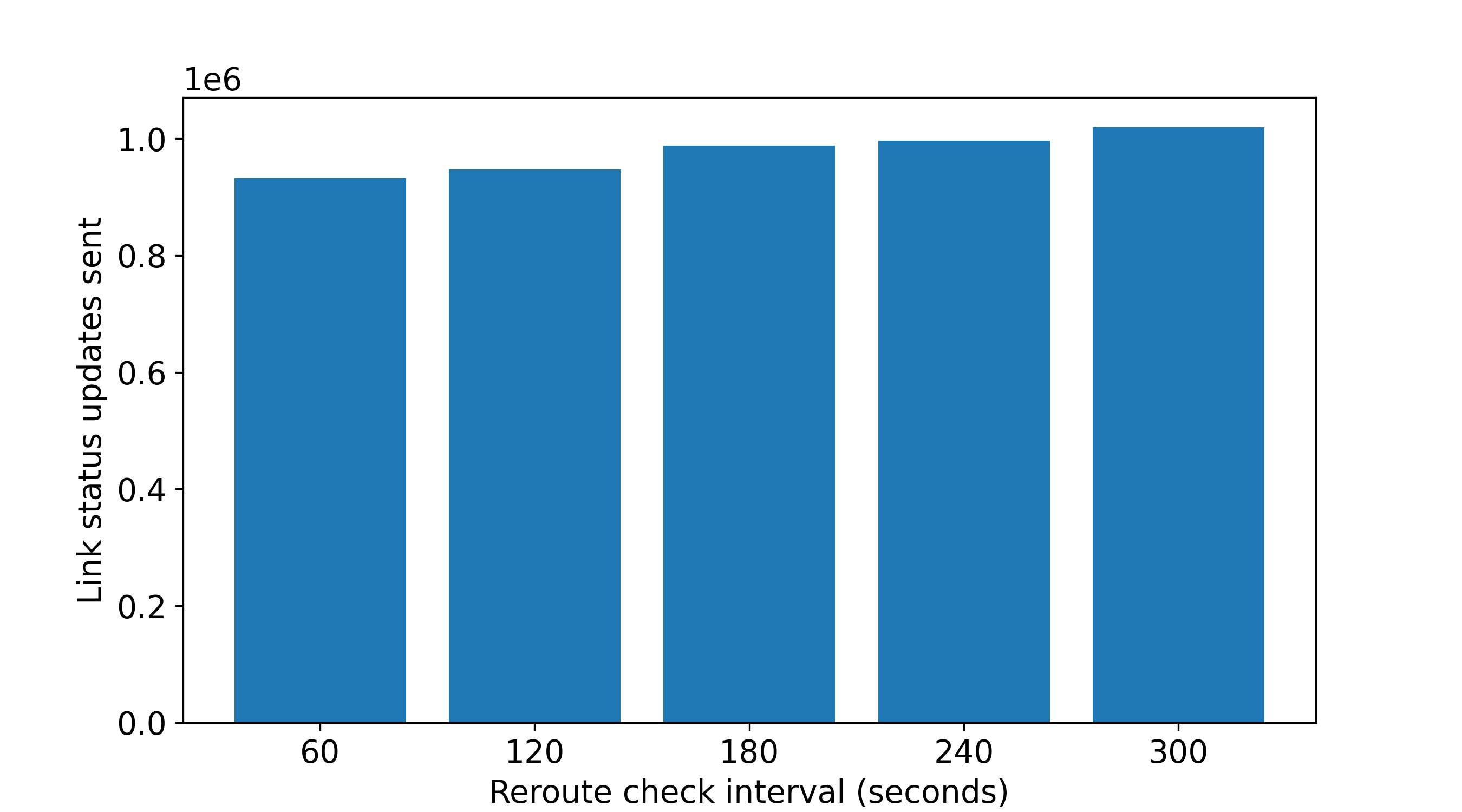}
        \caption{Link Status Updates Sent}
    \end{subfigure}
    \begin{subfigure}[b]{0.32\textwidth}
        \includegraphics[width=\textwidth]{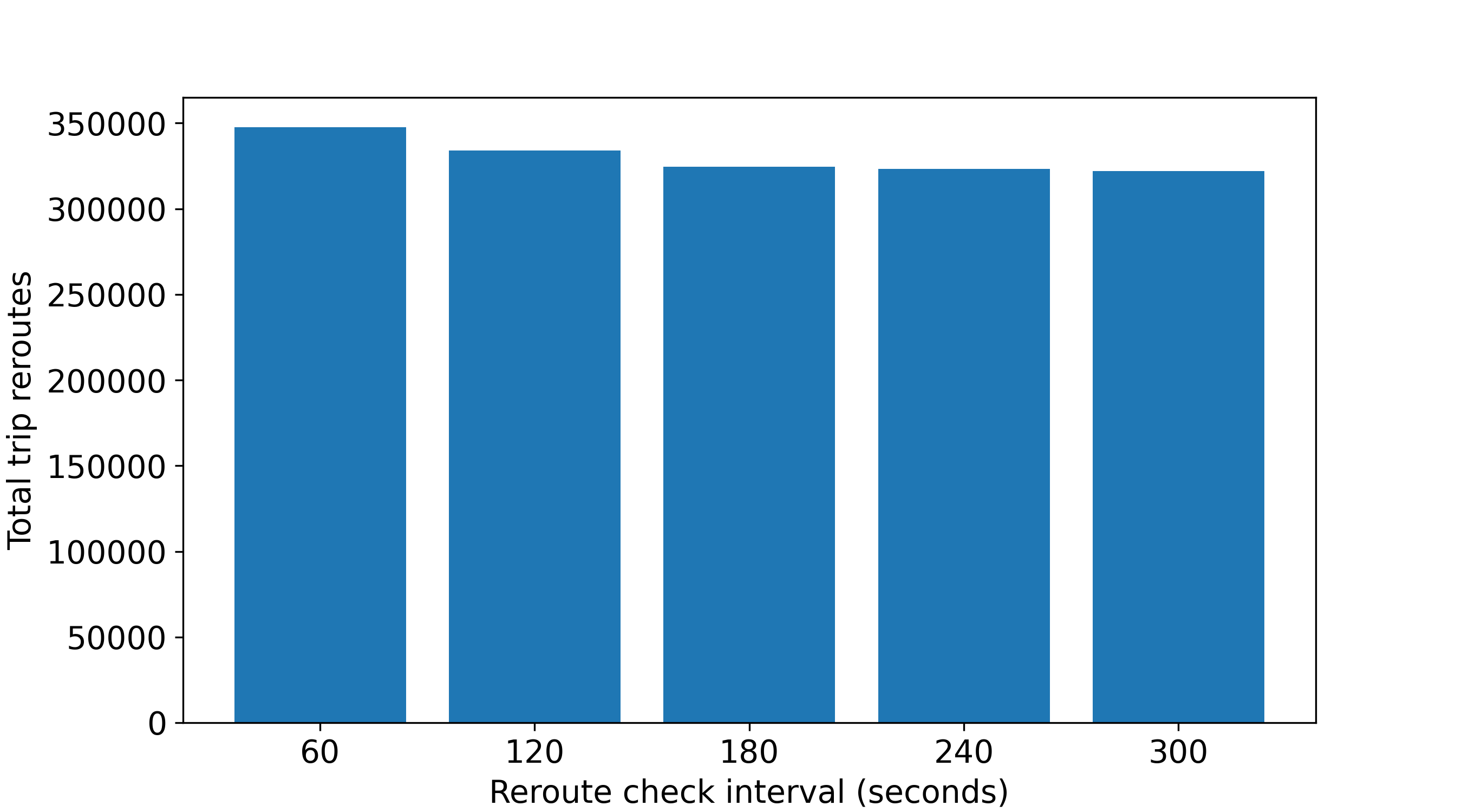}
        \caption{Trip Leg Reroutes}
    \end{subfigure}
    \begin{subfigure}[b]{0.32\textwidth}
        \includegraphics[width=\textwidth]{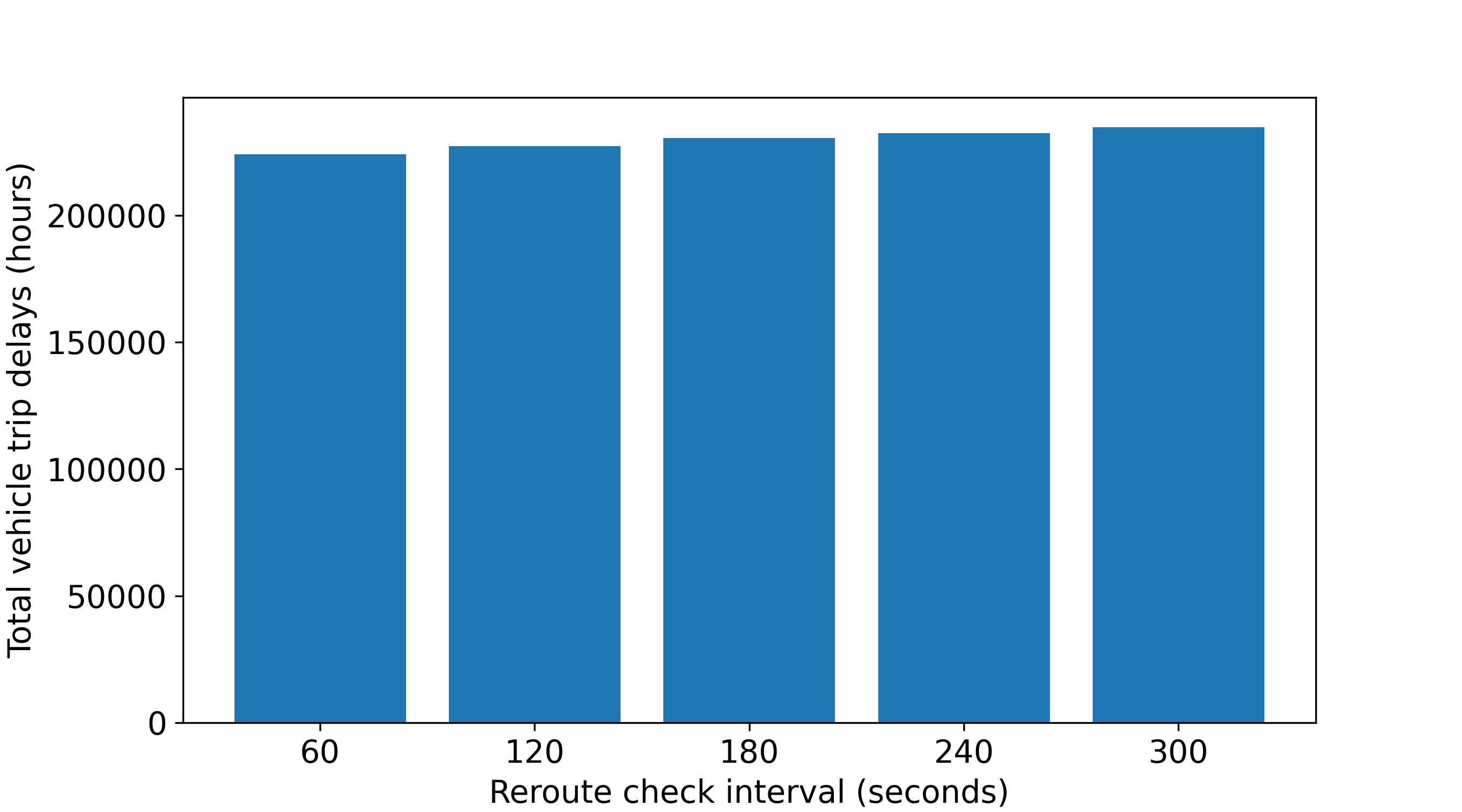}
        \caption{Total (System) Vehicle Delay}
    \end{subfigure}
    \caption{Parameter sensitivity sweep for rerouting check frequency. X-axis is the reroute check interval $t_\text{check}$.}
    \label{fig:check_sweep}
\end{figure*}

In the second parameter sweep shown in Figure~\ref{fig:check_sweep}, we varied the minimum time interval between a vehicle's reroute check queries: $t_\text{check}$ (on the x-axis).  As $t_\text{check}$ increases, vehicles check whether they should reroute \textit{less} frequently, but we observe that the total number of trip reroutes (b) declines only modestly.  This is because the vast majority of vehicles that reroute only have to check and switch to a better route \textbf{once}, and then it typically sticks to the new route (e.g., with 60\% rerouting penetration, 98.2\% of rerouted trips only reroute once during its journey).  Thus, increasing the time interval between checks mostly results in a small delay in the timing of a switch to a better route.  As a result, the impacts on the number of link status updates (a) and total system delay (c) are also relatively minor.

\begin{figure*}
    \centering
    \begin{subfigure}[b]{0.32\textwidth}
        \includegraphics[width=\textwidth]{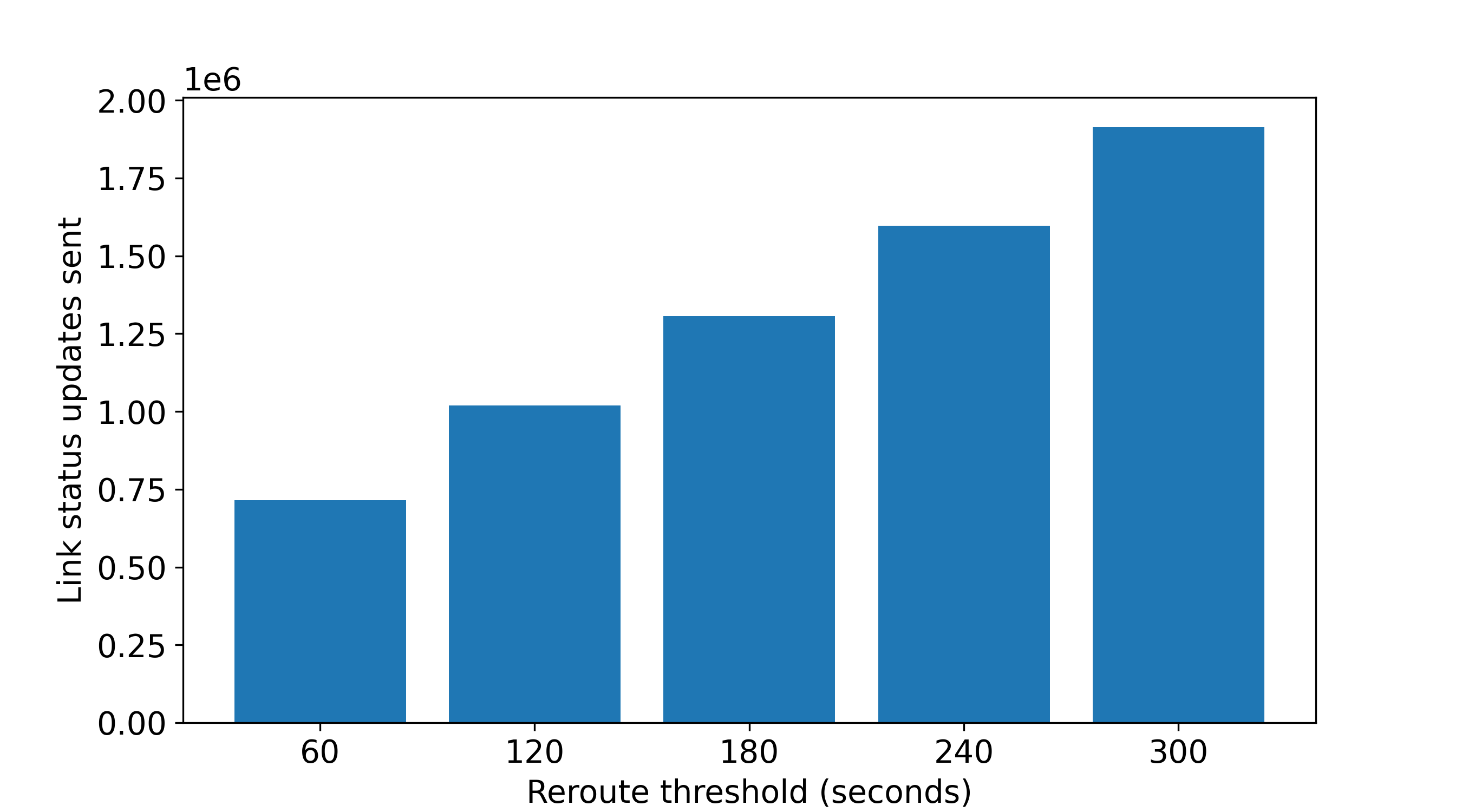}
        \caption{Link Status Updates Sent}
    \end{subfigure}
    \begin{subfigure}[b]{0.32\textwidth}
        \includegraphics[width=\textwidth]{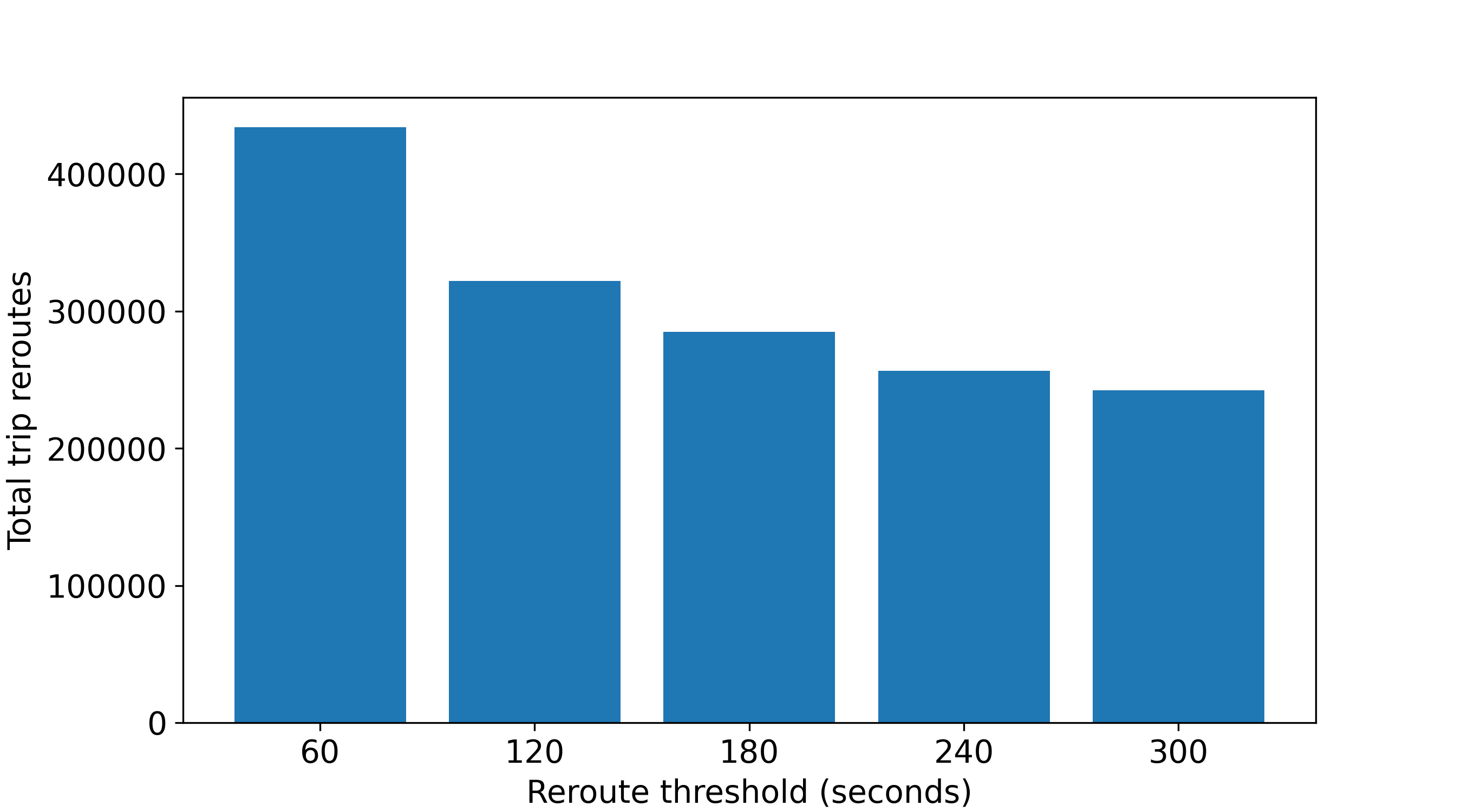}
        \caption{Trip Leg Reroutes}
    \end{subfigure}
    \begin{subfigure}[b]{0.32\textwidth}
        \includegraphics[width=\textwidth]{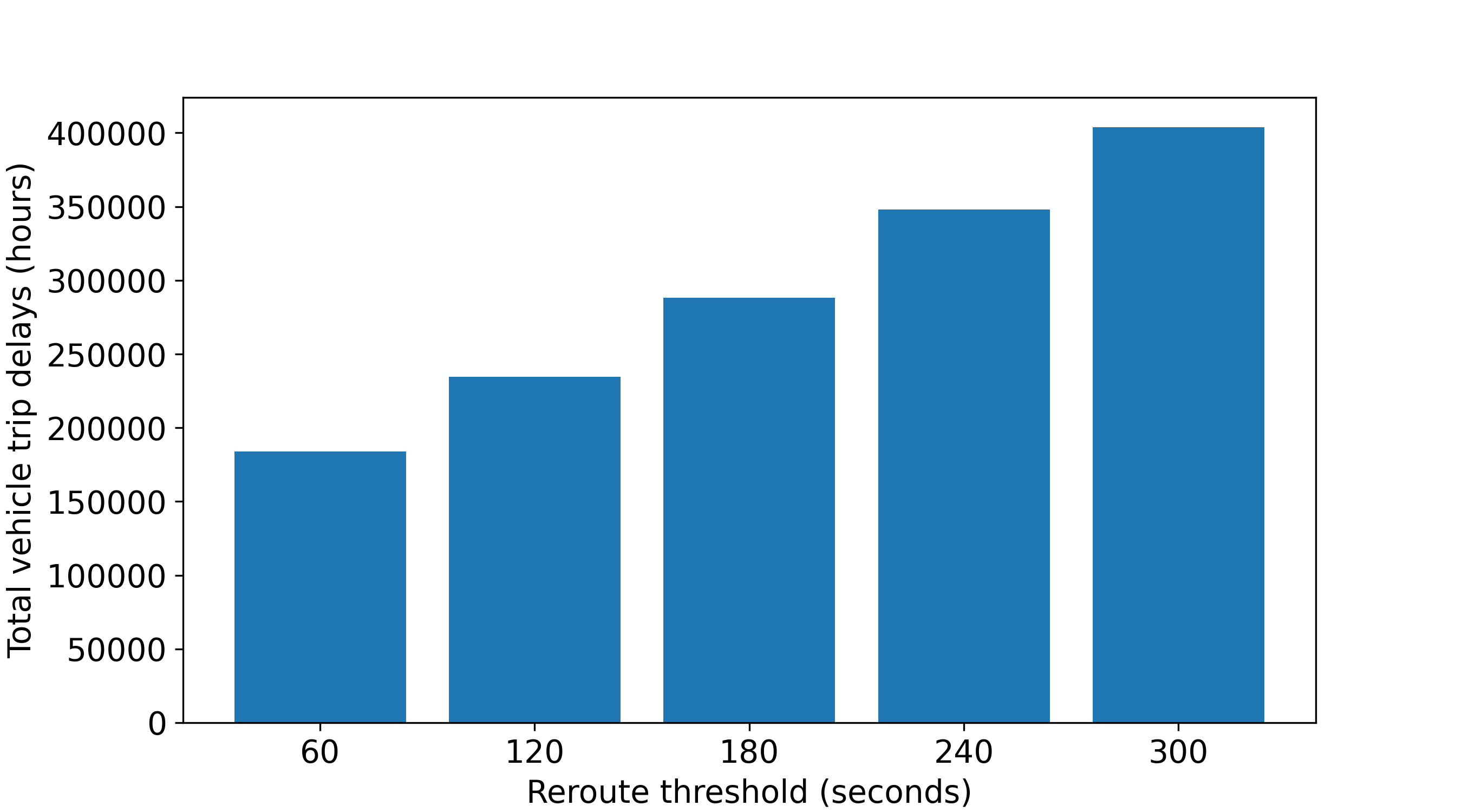}
        \caption{Total (System) Vehicle Delay}
    \end{subfigure}
    \caption{Parameter sensitivity sweep for rerouting delay and difference thresholds. X-axis is the absolute reroute delay threshold $t_\text{delay}$, where $r_\text{delay} = t_\text{delay} / 600$.}
    \label{fig:delay_sweep}
\end{figure*}

In the third parameter sweep shown in Figure~\ref{fig:delay_sweep}, we varied the delay thresholds for when a vehicle should reroute: $t_\text{delay}$ (on the x-axis) and set $r_\text{delay} = t_\text{delay} / 600$.  As $t_\text{delay}$ increases, the vehicle controllers are less aggressively rerouting vehicles, keeping them on their original paths until their current path congestion is higher and their best alternative route saves them more time.  This is directly seen in the middle graph as the total number of trip reroute requests decreases significantly as the thresholds increase.  The number of link status updates (a) increases due to more dynamic variation in congestion, as the controllers are less effective at balancing traffic among available alternatives.  The total system delay (c) increases since more vehicles are staying on less optimal routes, thus increasing their delay.  While the smallest threshold resulted in the best system efficiency, it is debatable whether using a very low delay threshold in the real world with human drivers is desirable because of the cognitive cost in asking a driver to alter their route.

For the experiments in the next section, we used the parameter values in Table~\ref{tab:reroute_params}, which provide a good balance between realism and improvement in system congestion.  The parameters with the largest effect on total system delay were the reroute threshold parameters ($t_\text{delay}$ and $r_\text{delay}$).  For these, we selected $t_\text{delay}$ equal to 2 minutes and $r_\text{delay}$ equal to 0.2 as a compromise between system efficiency and excessive rerouting.

\section{EXPERIMENTS}

\subsection{Experimental Methodology Description}

In order to demonstrate and evaluate our Mobiliti simulator, we followed the following steps to model and analyze road network traffic in the San Francisco Bay Area (the steps would be similar for other areas).  First, we obtained the necessary network, trip demand, and vehicle data inputs shown in Figure~\ref{fig:workflow}, left.  After obtaining the inputs, we run the Mobiliti simulator (Figure~\ref{fig:workflow}, center) for a variety of parameter configurations, including sweeping the dynamic rerouting penetration rate, the link status update threshold, the reroute check interval, and the reroute threshold.  The simulator outputs include link-level metrics such as vehicle counts, average speeds, and congestion delay series; leg-level metrics such as trip times and delays; and rerouting information such as when and where rerouting occurs.  We then process the simulator logs and outputs (Figure~\ref{fig:workflow}, right) using a combination of software tools (such as Python) to conduct the various analyses presented in the following subsections.

The road network model was derived from a HERE Technologies~\cite{here_tech} map consisting of 454,651 nodes and 1,008,959 links spanning from Santa Rosa, Napa, and Vacaville to the north, San Jose to the south, and Oakland, Hayward, Fremont, and Livermore to the east (see Figure~\ref{fig:partitions}). 
%The map information is transformed into a different representation in order to integrate into the Mobiliti platform. 
While the link actor model currently supports the signal timing mechanism described in Section~\ref{sec:link_model}, the following experiments were run without detailed signal behavior due to lack of comprehensive, accurate location and timing data for all of the signals in the Bay Area.  However, the link capacity properties in the input road network already take into account the presence of signals and are used in each link actor's Congestion Delay Model (see Section~\ref{sec:link_model}) to compute vehicle traversal times, thus the simulator slows vehicles according to those flow capacity values compared to having no signals.

\begin{figure}[h]
    \centering
    \includegraphics[width=0.43\textwidth]{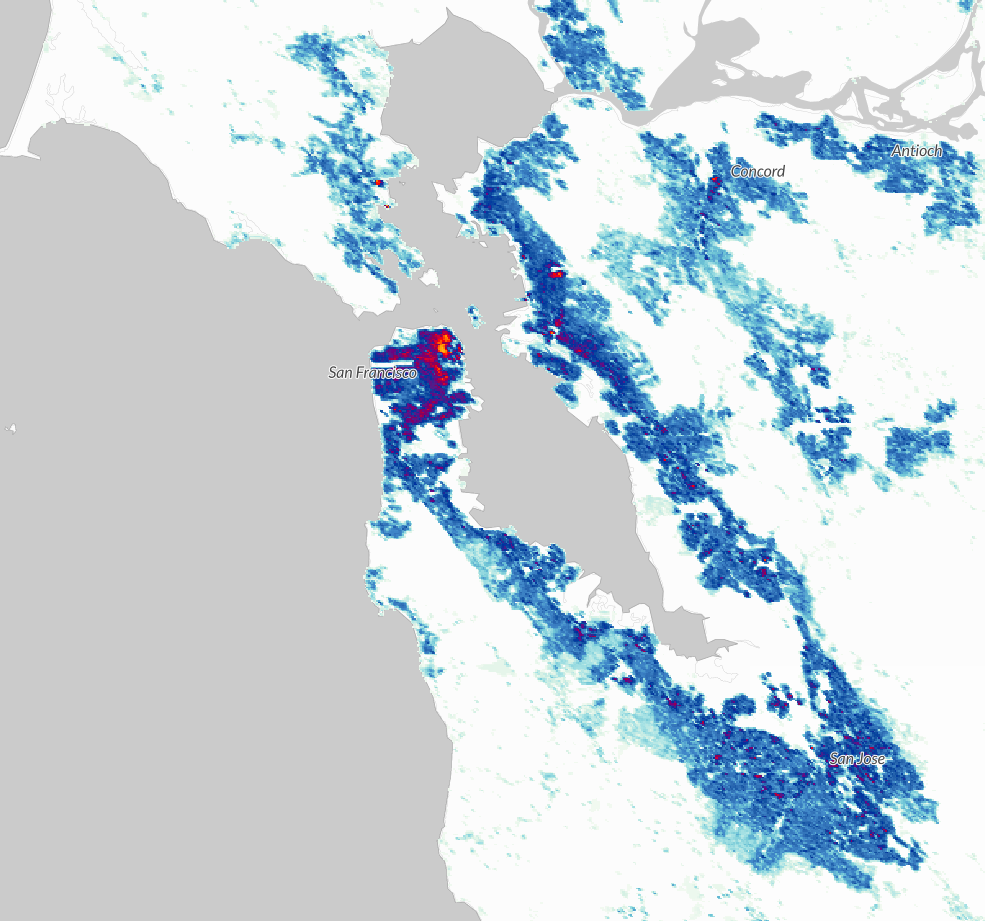}
    \caption{GHS density map~\cite{ghs} showing where people live as a function of geographical location. Mobiliti uses this information coupled with a TAZ-to-TAZ demand model to compute specific node origins and node destinations for each trip leg.  The raster resolution of the GHS data is 250 meters by 250 meters. Red shows high population density, followed by dark blue with medium density and lighter blues with low density. }
    \label{fig:pop_density}
\end{figure}

The trip demand is initialized from an input file with 19 million trip legs (origin/destination pairs) based on disaggregate simulated trip records from the San Francisco County Transportation Authority (SFCTA) SF CHAMP 6.1 model~\cite{sf-champ6.1}. Each trip leg is specified with origin and destination travel analysis zones (chosen from 40,000 micro-analysis zones) and a start time. Since our simulator models each individual vehicle traversing from link to link at discrete times, we chose specific origin and destination nodes within the given TAZs. We weighted each node by its nearby population density derived from the Global Human Settlement database~\cite{ghs} to avoid choosing nodes that are in very sparsely populated regions of the map, which would unrealistically send traffic to remote areas.  We also avoided selecting freeway or ramp nodes as origins or destinations.  Figure~\ref{fig:pop_density} shows the population density map we used for initializing our simulated trip legs.  Note that the coarse granularity of the heat map is a result of the resolution of the provided data set (250 meters).

\begin{figure}
    \centering
    \begin{subfigure}[b]{.45\textwidth}
        \includegraphics[width=\textwidth]{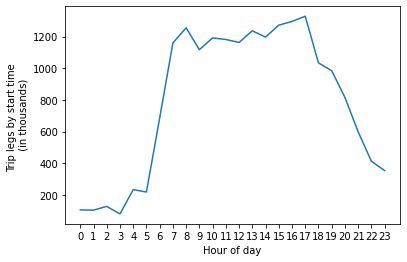}
        \label{fig:trip_legs}
        \caption{(a) Total trip legs by start time}
    \end{subfigure}
    \begin{subfigure}[b]{.45\textwidth}
        \includegraphics[width=\textwidth]{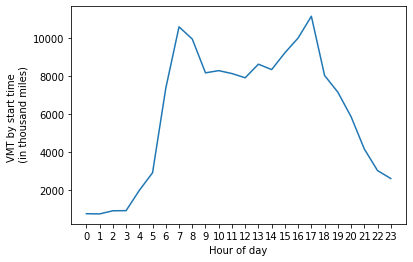}
        \label{fig:trip_distance}
        \caption{(b) VMT (total trip distance) by start time}
    \end{subfigure}

    \caption{These ﬁgures give a temporal proﬁle of how demand evolves through the simulated day. Figure (a) shows the total number of trip legs starting at diﬀerent times of the day. Figure (b) shows the VMT (total trip distance) summed over trips that start at different times of day.}
    \label{fig:legs_by_time}
\end{figure}

Figure~\ref{fig:legs_by_time}a shows the temporal profile of the SFCTA demand model's trip legs during the simulated model day.  Our simulation runs a single model day, and the figure shows on the y-axis the number of trips that start in each hour of the day.  The number of trip legs per hour varies from very low in the early morning hours to very high during late afternoon rush hour.  Since trip lengths can vary considerably, by weighing each trip by its length, Figure~\ref{fig:legs_by_time}b shows the VMT (total trip distance) by start time, which is defined as the sum of the total trip distance over all trips that start in each hour of the day.  This figure illustrates the time-varying magnitude of the total load on the road network.

\subsection{Computational Performance}

This section details the design and analysis of the computational performance of the Mobiliti simulator.  For our experiments, we ran Mobiliti on the Cori supercomputer~\cite{cori}. Adding dynamic rerouting adds significant computational cost to the simulation compared to the original statically routed case. This is mainly due to the addition of shortest path route calculations by VehicleController actors during RerouteCheck events, and also due to the additional processing of LinkStatusUpdate events.  Not only does the cost of calculating new routes directly add to the simulation time, but the irregularity of congestion also causes load imbalance of compute workload across the simulator's parallel compute cores.
We mitigate this effect by using an efficient routing algorithm (Customizable Contraction Hierarchies~\cite{geisberger12-ch,dibbelt14-cch}) via the RoutingKit~\cite{routingkit} library. RoutingKit’s Customizable Contraction Hierarchies library employs relatively costly preprocessing steps (i.e. partial customizations) to enable very efficient subsequent routing queries. This strategy is ideal for computing many routes in a network with static weights, since the cost of the preprocessing step can be amortized across all of the subsequent queries.

However, in a dynamic rerouting simulation the link weights are constantly changing due to congestion varying over the course of the simulated day. In order to support dynamically changing link weights, we designed the VehicleController to keep track of which links have updated their travel time since the previous customization, and then only re-customize the contraction hierarchy when a RerouteCheck query is received. This method of batching LinkStatusUpdates avoids excessive re-customizations every time an update is received.  Furthermore, if no congestion update is received during a sequence of RerouteCheck queries, they can all use the same customization and thus be serviced very efficiently.

\begin{figure}[h]
    \centering
    \begin{subfigure}{.48\textwidth}
        \includegraphics[width=\textwidth]{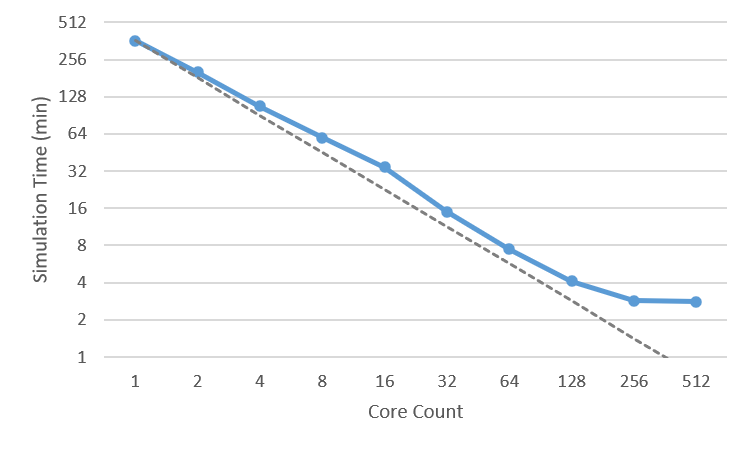}
        \caption{}
        \label{fig:scaling_perf}
    \end{subfigure}
    \begin{subfigure}{.48\textwidth}
        \includegraphics[width=\textwidth]{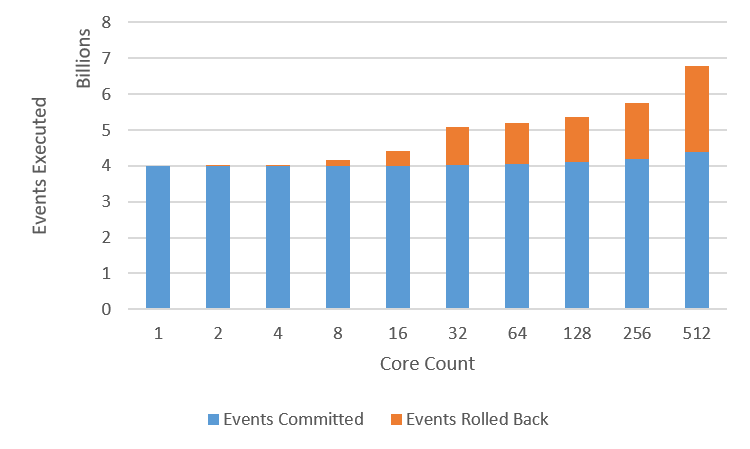}
        \caption{}
        \label{fig:scaling_events}
    \end{subfigure}
    \caption{(a) Mobiliti simulation times (excluding program initialization) as we vary the core count from 1 to 512 cores of the Cori-Haswell computer at NERSC~\cite{cori}. All times shown are for simulating 19 million trip legs with 50 percent dynamic rerouting penetration on the whole San Francisco Bay Area road network with 0.5 million nodes and 1 million links.  When simulating a normal model day, execution time is decreased from over six hours on one core to less than three minutes on 512 cores running in parallel. (b) Mobiliti total events executed as the core count is varied from 1 to 512 cores.  Executed events are either committed (successful) or rolled back (unsuccessful).}
\end{figure}

Figure~\ref{fig:scaling_perf} shows the parallel scaling performance of Mobiliti when simulating a full normal model day with 19 million trip legs over the San Francisco Bay 0.5 million nodes and 1 million links with 50 percent dynamic rerouting penetration. The dashed line represents theoretical perfect linear scaling performance.  As we increase the core count from 1 to 512, the simulation execution time (excluding program initialization) is reduced from more than six hours to less than three minutes, corresponding to a 130x speed up versus serial execution.

For the shared memory runs (16 cores or fewer), we use a single process, multi-threaded configuration (utilizing one thread per core). Using shared memory avoids the overheads of inter-process communication, but limits execution to a single node. For the distributed memory runs (32 cores or more), we use a multi-process, multi-threaded configuration with hyper-threading (two threads per core), which enables executing on multiple nodes, better data locality in the memory subsystem due to non-uniform memory access \cite{majo2011memory}, and higher core utilization.  Furthermore, strong scaling (increasing core count for a fixed problem size) yields the benefit of reducing the active working set size for each core, enabling better data re-use and on-chip cache utilization.  Figure~\ref{fig:scaling_events} shows the total number of events committed and rolled back for each of the runs in the scaling study.   For the single core run, there is no misspeculation because events are trivially executed in increasing timestamp order.  As the number of cores increases, commit efficiency (events committed / total events executed) levels off around 90 percent for the shared memory runs and ranges from 79 to 65 percent for distributed memory runs.

We note that performance improvements from increasing parallelism level off at 512 cores and suspect the scalability could be further improved for even higher core counts by tuning the parallel distribution of the dynamic re-routing workload to refine the computational load balance and reduce mis-speculation overhead in the simulator.  A more in-depth parallel performance study of the simulator will be explored in future work. 

\subsection{Impact of dynamic rerouting on system metrics}
\begin{figure}
    \centering
    \begin{subfigure}[b]{.45\textwidth}
        \includegraphics[width=\textwidth]{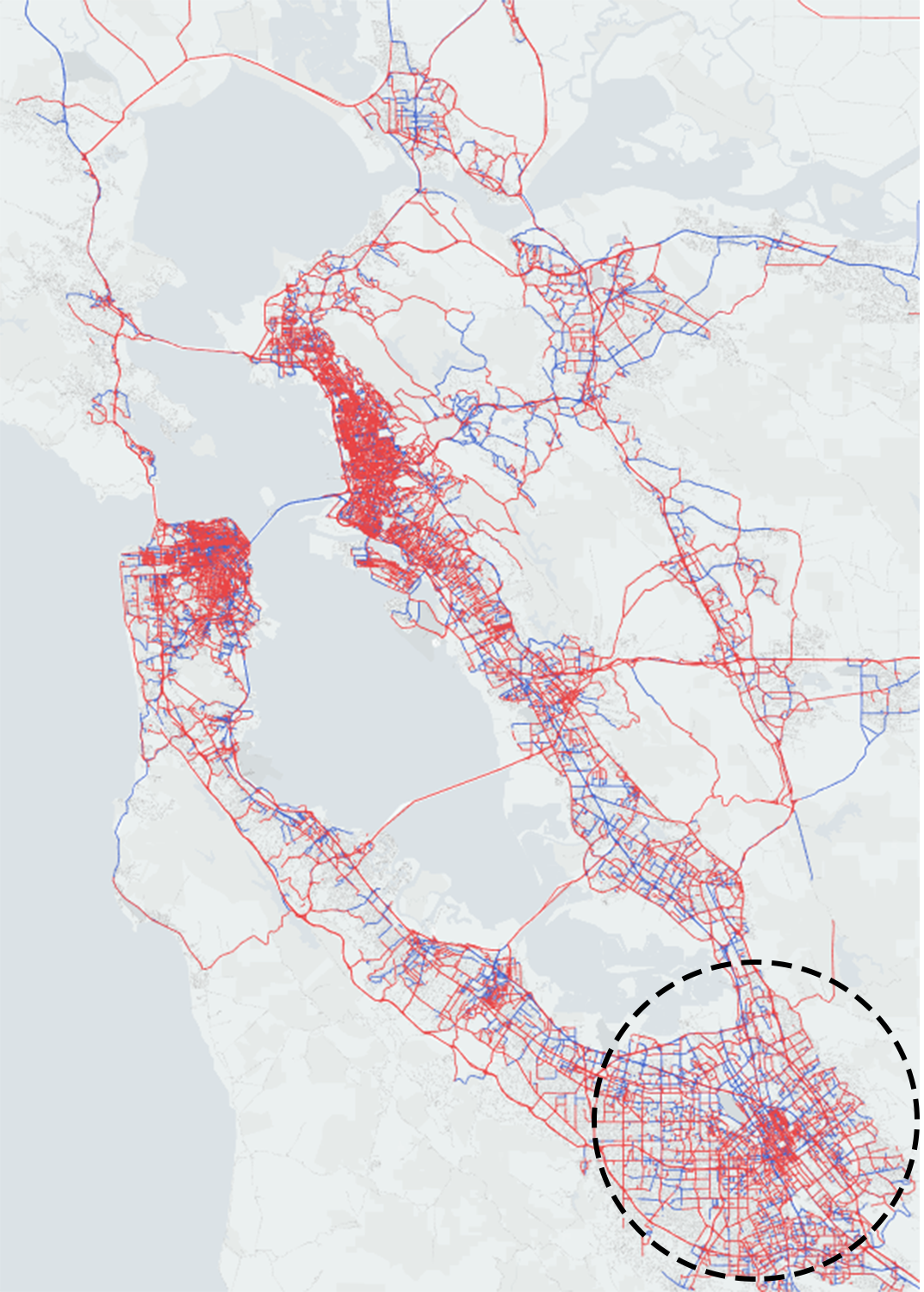}
        \caption{(a) Bay Area}
    \end{subfigure}
    \begin{subfigure}[b]{.45\textwidth}
        \includegraphics[width=\textwidth]{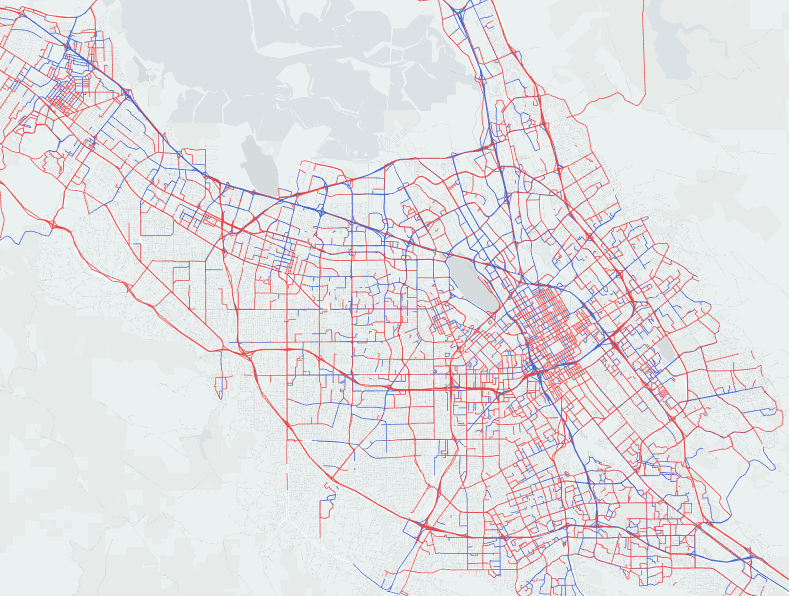}
        \caption{(b) San Jose Region}
    \end{subfigure}
    \caption{System-wide impact on number of vehicle link traversals from enabling dynamic rerouting (Baseline versus 100\% penetration). Red or blue represent a increase or decrease in vehicle traversal count respectively for a day.}
    \label{fig:wholebay}
\end{figure}

To understand the effect of rerouting penetration, we enabled dynamic rerouting for varying percentages of vehicle trip legs for the entire Bay Area.  We chose to study a range of penetration rates from 0\% to 100\% at 10\% increments. Figure~\ref{fig:wholebay}  illustrates the impact of enabling dynamic rerouting for 100\% vehicle trips for the entire metropolitan-scale system simulated. The difference between baseline and 100\% rerouting case is that the former uses static shortest path routes based on free speed traversal times, whereas the latter uses dynamic routes computed by the vehicle controllers based on their knowledge of the current traffic congestion patterns.  In Figure~\ref{fig:wholebay}, blue links handle a lesser number of vehicle traversals when 100\% dynamic rerouting is enabled, while the red links handle a greater number. The figure shows how the traffic is rerouted away from certain links to reduce congestion (blue), while other links end up handling higher traffic (red). 
 
\begin{figure}[h]
    \centering
    \includegraphics[width=0.5\textwidth]{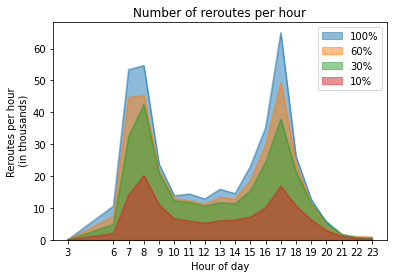}
    \caption{The figure give a temporal profile of the rate
of dynamic vehicle reroutes for a day. Sharp increases in number of reroutes can be seen in the morning and evening peak hours.}
    \label{fig:reroutes_per_hour}
\end{figure}

Figure ~\ref{fig:reroutes_per_hour} shows how the reroutes are temporally distributed throughout the simulation day, illustrating that almost 80\% of the rerouting occurs during the morning and evening rush hours when the demand is the highest. The distribution of reroutes is heavily influenced by the temporal distribution of trip legs in the demand model input (Figure~\ref{fig:legs_by_time}). As can be seen in Figure~\ref{fig:legs_by_time}b, there is a peak in the VMT in the morning (7am to 10 am) and the evening rush hours (3:30 pm to 6:30 pm). Because the level of demand during the rush hours is the highest, we see corresponding peaks in the number of reroutes. Further, it needs to be noted that penetration rate indicates the number of trip legs that are allowed to reroute, but not all reroutable trip legs do actually reroute. Table \ref{tab:actually_rerouted} indicates the percentage of trip legs that actually got rerouted as a percentage of allowed reroutable legs. It can be seen that at higher penetration rates the percentage decreases even though the actual number of reroutes is higher, as not many trip legs engage in any relevant congestion and hence do not reroute.  Furthermore, among the trips that do reroute, the number of \textit{reroutes per trip} remains small, with 99.9\% of trips rerouting three times or fewer in the 100\% penetration scenario.

\begin{table}[h]
\small
\centering
\caption{Share of Rerouted Trip Legs }
\label{tab:actually_rerouted}  
\begin{tabular}{cccc} 
\hline
\textbf{Penetration rate} &  \textbf{ Trip legs reroutable} & \textbf{ Trip legs rerouted} & \textbf{Percentage rerouted} \\
{(\%)} & {(in thousands)}  & {(in thousands)} & {(\%)}\\
\hline
\noalign{\smallskip}
10 & 1,900 & 130 & 7 \\ 
20 & 3,800 & 221 & 6 \\ 
30 & 5,700 & 274 & 5 \\ 
40 & 7,600 & 291 & 4 \\ 
50 & 9,500 & 301 & 3 \\ 
60 & 11,400 & 317 & 3\\ 
70 & 13,300 & 329 & 2\\ 
80 & 15,200 & 340 &  2\\ 
90 & 17,100 & 355 & 2\\ 
100 & 19,000 & 374 & 2\\ 
\noalign{\smallskip}
\hline
\end{tabular}
\end{table}

\begin{figure}
    \centering
    {
        \includegraphics[width=0.5\textwidth]{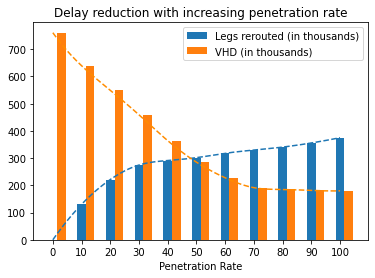}
    }
    \caption{System wide vehicle hours of delay (VHD) and the number of trip legs rerouted from varying the dynamic rerouting penetration rates. As the penetration rate increases, the overall system delay reduces. If we look at the elbow of the VHD curve, after 70\% penetration rate the returns start diminishing.}
    \label{fig:systemwide}
\end{figure}

\begin{figure}
    \centering
    \begin{subfigure}{.5\textwidth}
        \includegraphics[width=\textwidth]{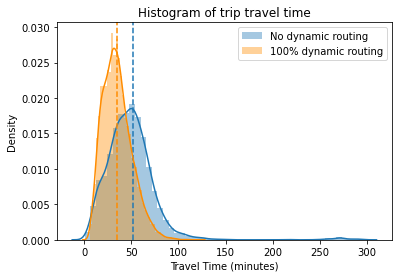}
        \caption{Distribution of the travel times for trip legs.  The mean travel time per trip leg with 100\% dynamic routing is 35 minutes and without is 51 minutes as showed by the corresponding orange and blue vertical lines.}
        \label{fig:rerouting_timesaved}
    \end{subfigure}
    \begin{subfigure}{.45\textwidth}
        \includegraphics[width=\textwidth]{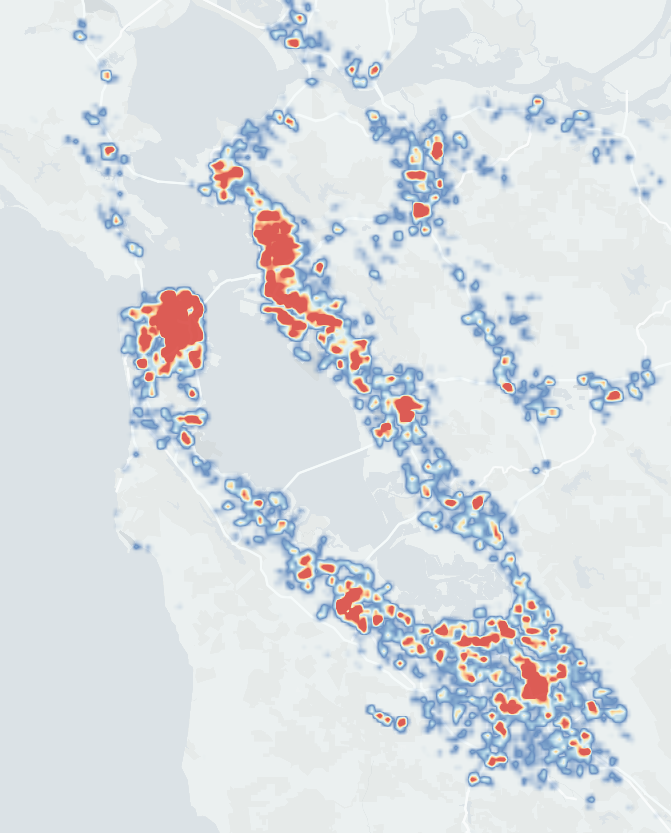}
        \caption{Heat map of areas with highest congestion (v/c over 0.75) with  increased vehicle counts due to dynamic rerouting  in the morning rush hour for functional class 5 links. Red indicates high congestion and blue represents low congestion.}
        \label{fig:fc5heatmap}
    \end{subfigure}
    \caption{}
\end{figure}

Using the delay and fuel model described in our previous work~\cite{chan_mobiliti_2018}, we are able to make impact estimates for dynamic rerouting penetration rates. Figure~\ref{fig:systemwide} exhibits the system level vehicle hours of delay (VHD) and number of reroutes for different penetration levels and Table~\ref{tab:vmt} shows the system level vehicle miles travelled (VMT) and fuel consumption resulting from the dynamic rerouting. As the penetration rate increases, the delay reduces without any significant change in VMT. The minimum system delay is obtained with 100\% penetration rate. However, if we look at the ``knee of the curve'' for VHD, the return starts diminishing after 70\% penetration. On average, a rerouted trip saves 16 minutes in travel time with no additional trip distance with 100\% penetration rate compared to baseline as shown in Figure \ref{fig:rerouting_timesaved}. 

% We note that the maximum benefit of dynamic rerouting in terms of reducing the total system delay is obtained with 75\% penetration rate. There is an increase of 1000 hours of delay with 100\% penetration.  A plausible explanation for this effect is the overall reduction of system efficiency when everyone tries to change routes based on the current traffic conditions, leading to an over-correction and worse observed congestion.  Thus, the largest system benefit from dynamic routing devices occurs when only a subset of drivers employs them.

% In the case where over-correction becomes an issue due to a high rerouting penetration rate, the vehicle controllers can adjust their behavior to lower the effective penetration rate closer to optimal.  Specifically, if the vehicle controllers select some fraction of vehicle requests to keep their existing route, then the effective penetration rate of the population is reduced by that fraction.  In this way, the vehicle controllers can close the gap to the optimal penetration rate by directing some vehicles to utilize their current routes instead of steering all traffic to the (currently) non-congested routes.

\begin{table}[h]
\small
\centering
\caption{VMT and Fuel Consumption}
\label{tab:vmt}      
\begin{tabular}{ccc}
\hline
\textbf{Penetration rate} & \textbf{VMT} & \textbf{Fuel}\\
 & {(in thousand miles)} & {(in thousand gallons)}\\
\hline
\noalign{\smallskip}
0\% & 146,847  & 5,906\\
10\% & 146,783 & 5,903\\ 
20\% & 146,707 & 5,899\\ 
30\% & 146,652 & 5,894\\
40\% & 146,605  & 5,891\\ 
50\% & 146,572  & 5,888\\
60\% & 146,546  & 5,886\\ 
70\% & 146,537 & 5,884\\ 
80\% & 146,517  & 5,883\\ 
90\% & 146,505 & 5,882\\ 
100\% & 146,490 & 5,882 \\ 
\noalign{\smallskip}
\hline
\end{tabular}
\end{table}

We observe that dynamic rerouting effectively rearranges the vehicle flows from high utilization highways and arterials to low utilization neighborhood links to reduce the overall system delay.  We analyze these effects by investigating the rearrangement of traffic flow by functional classification of links.  We maintain the definitions of functional class roads as defined by HERE Technologies~\cite{here_tech}. Specifically, functional classes are hierarchical classification of roads according to the speed, importance and connectivity of the road. A road can be one of five functional classes defined in Table~\ref{tab:functionalclasses}.

\begin{table}[h]
\small
\centering
\caption{Functional Road Classes}
\label{tab:functionalclasses}
\begin{tabular}{cc} 
\hline \textbf{Functional class} & \textbf{ Definition}\\
\hline
\noalign{\smallskip}
1 & Allows high volume, maximum speed traffic movement \\
2 & Allows high volume, high speed traffic movement \\
3 &  Provides high volume of traffic movement \\
4 & Provides high volume of traffic movement at moderate speeds between neighborhoods \\
5 & Local roads with volume and traffic movement below the level of any functional class\\
\noalign{\smallskip}
\hline
\end{tabular}
\end{table}

By examining the traffic flow by functional classes with 100\% dynamic rerouting penetration, we observe that traffic shifts from FC 2 and 3 to FC 4 and 5. This reduces the delay on highways significantly while increasing traffic on FC 5 in the morning and evening peaks. It is also interesting to note that the increase in traffic volume on FC 5 does not always cause congestion in those links as many links do not reach congested levels with the increased flow. Specifically, 7000 kilometers of FC 5 links received additional traffic flow with 100\% dynamic rerouting, of which 75\% received fewer than 6 additional vehicles during the morning peak.

Of the FC 5 links with increased traffic volume, 440 kilometers are congested with a volume over capacity ratio higher than 0.75. Spatial analysis of the congestion shows the cities of San Francisco, San Jose, Berkeley, Oakland and Fremont are the most affected by the increased traffic in the local roads (Figure ~\ref{fig:fc5heatmap}). The local roads (FC 5) in these cities has a mean increase of 70 vehicles during the morning peak.  Finally, of the 440 kilometers of congested road network in the morning peak, 110 kilometers reflects \textit{new} congestion created due to dynamic rerouting on the local roads.  These roads would arguably be some of the most negatively impacted areas by high dynamic rerouting penetrations.

Finally, we present the results of user equilibrium traffic assignment (UE) in Table \ref{tab:uet} for comparison using the methods we describe in \cite{DBLP:journals/corr/abs-2104-12911}. Comparing with the system metrics for dynamic rerouting with 100\% penetration rate, we can see that VHD is nearly half, fuel consumption is slightly higher and VMT slightly lower in user equilibrium.  Since the user equilibrium is a steady-state solution computed through iterative optimization, it results in routes with lower delays than the more reactive dynamic rerouting approach.  However, in reality, the user equilibrium state in never actually achieved and hence congestion is underestimated in the user equilibrium case.

\begin{table}[h]
\small
\centering
\caption{User Equilibrium System Metrics}
\label{tab:uet}      
\begin{tabular}{cccc}
\hline
\textbf{} & \textbf{VMT} & \textbf{VHD} & \textbf{Fuel}\\
 & {(in thousand miles)} & {(in thousand hours)} & {(in thousand gallons)}\\
\hline
\noalign{\smallskip}
User Equilibrium & 146,051 & 90 & 5,915 \\
\noalign{\smallskip}
\hline
\end{tabular}
\end{table}

\subsection{Validation}
Validation was performed for the simulation runs with different penetration rates to test the effectiveness of representing the real world traffic environment. We conducted a three stage validation procedure using multiple data sources. Our results show that the simulation with 60\% dynamic rerouting is the closest to representing the real world traffic. For brevity, we have only included the validation for this penetration rate here. Our results are also consistent with multiple surveys stating that the percentage of Americans having smartphones who uses online maps or navigation services daily ranges from 55\% to 65\% \cite{noauthor_people_nodate, noauthor_maps_nodate, noauthor_popularity_nodate}. 

\begin{figure}[ht]
  \includegraphics[width=.3\columnwidth]
    {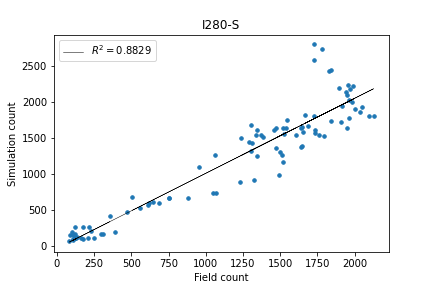}\hfill
  \includegraphics[width=.3\columnwidth]
    {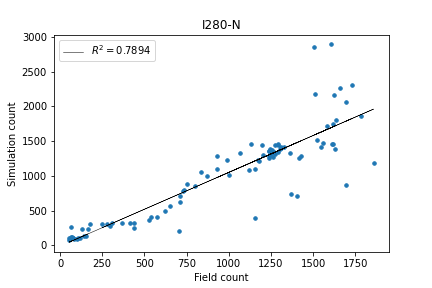}\hfill
  \includegraphics[width=.3\columnwidth]
    {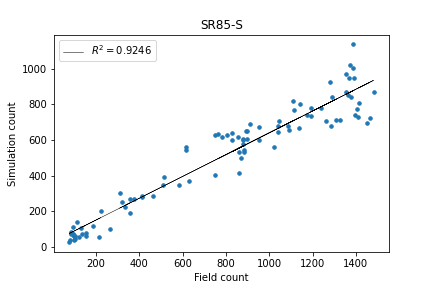}
  \includegraphics[width=.3\columnwidth]
    {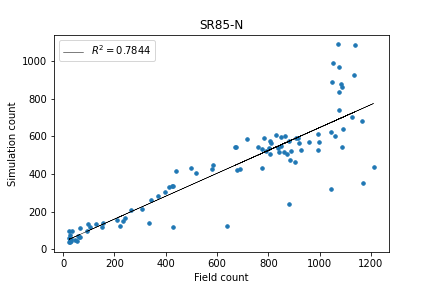}
  \includegraphics[width=.3\columnwidth]
    {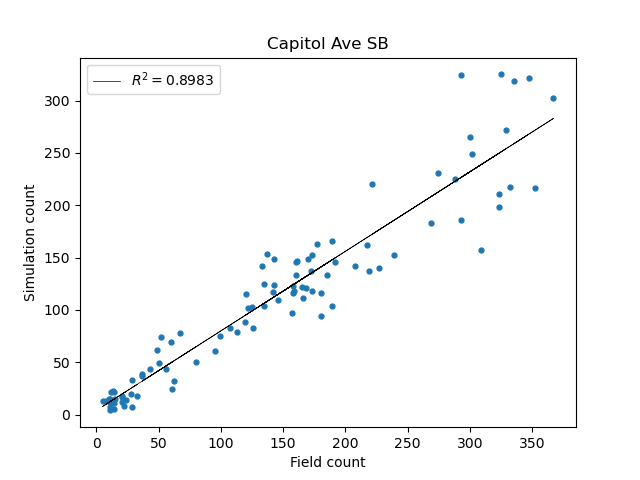}
  \includegraphics[width=.3\columnwidth]
    {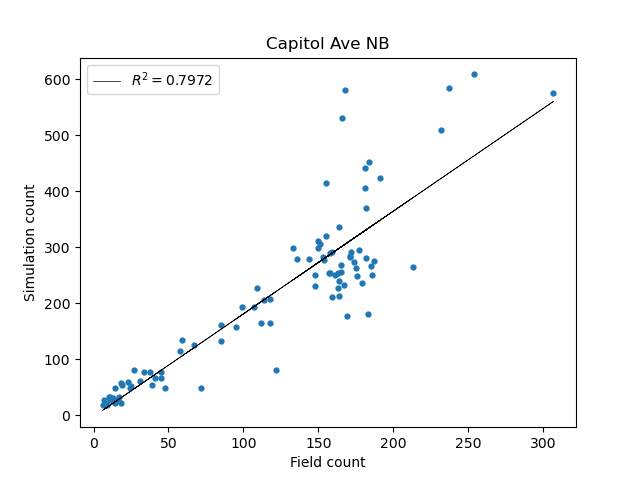}
  \includegraphics[width=.3\columnwidth]
    {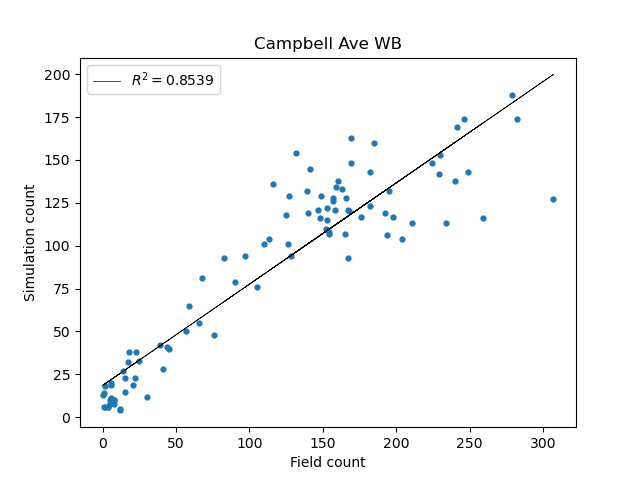}
  \includegraphics[width=.3\columnwidth]
    {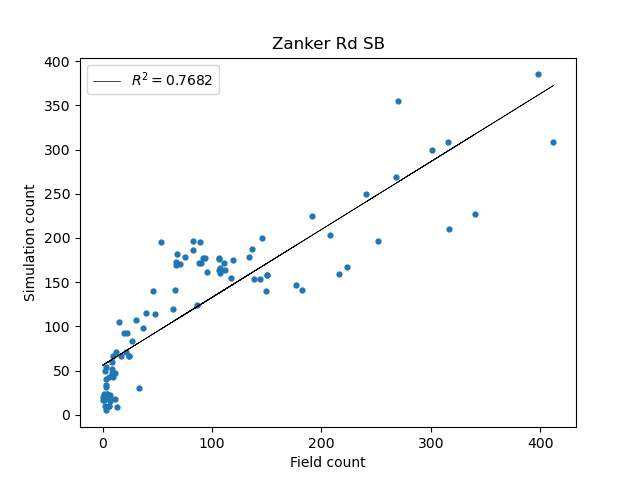}
  \caption{Validation of traffic counts in 15 minute increments for links in different functional classes. All links have satisfactory R$^{2}$ of greater than 0.7.}
  \label{fig:stage1a}
\end{figure}

 Stage 1 of the validation procedure involves comparing the traffic flows or counts between the simulation and real world. This includes checking a) traffic counts for eight corridor links in San Jose city, b) average daily traffic counts (ADT) for four main bridges in the Bay Area, and c) traffic flows for all major highways in the Bay area. The traffic count for each link was compared against the field data for the entire day in 15 minutes increments. The field data for city roads and highways were  collected from the city of San Jose and Caltrans PeMS website \cite{noauthor_traffic_nodate} respectively for the year 2019. Each corridor provided information regarding traffic volumes by time of the day and direction. For PeMS data, since it is prone to measurement error, data from multiple weekdays in April and May 2019 were averaged to get a typical day value. 
 %For San Jose city, data we received was for one day in October 2019. 
 A coefficient of determination (R$^2$) of 0.7 which is typically used as a satisfactory criterion for link count checks is used as the threshold.  The Figure \ref{fig:stage1a} shows R$^{2}$ values for the eight corridors under consideration. The modeled corridors indicate a close match with the field data with the lowest R$^2$ value observed being 0.76 for Zanker road.
 
 Additionally, ADT for four main bridges in the Bay Area in both directions is shown in Table~\ref{tab:stage1b}. The field data was obtained from the Caltrans website \cite{noauthor_traffic_nodate} for the year 2019. The target error was $\pm25\%$, and seven out of the eight links met this criterion.  We believe that the high relative error for the Golden Gate Bridge NB count is due to the Caltrans field count not being an accurate representation of the actual bridge count.  The discrepancy is due to the sensor placement \textbf{after} a major exit, which results in a significant percentage of bridge traffic not being counted by the sensor. 
 
 %Simulation counts for the Bay Bridge EB were higher than the field count, which is likely due to sensor location, lack of a toll plaza in the simulation, or a discrepancy between real-world demand and the synthetic demand model.
 
\begin{table}[ht]
\small
\centering
\caption{ADT Check}
\label{tab:stage1b}      
\begin{tabular}{ccccc}
\hline \textbf{Sl No} & \textbf{Bridge} & \textbf{Field Count} & \textbf{Simulation Count} & \textbf{Relative Error (\%)} \\
\hline
\noalign{\smallskip}
1 & I-580 Richmond San Rafael Bridge EB & 56182 & 51551 & -8\% \\
2 & I-580 Richmond San Rafael Bridge WB & 41597 & 52131 &  25\% \\
3 & I-80 Bay Bridge EB & 132000 & 148105 & 12\% \\
4 & I-80 Bay Bridge WB & 131861 & 139993 & 6\% \\
5 & US-101 Golden Gate Bridge NB & 32212$^*$ & 63730 & 98\% \\
6 & US-101 Golden Gate Bridge SB & 74526 & 70020 & -6\% \\
7 & CA-92 San Mateo Bridge EB & 56510 & 53684 & -5\% \\
8 & CA-92 San Mateo Bridge WB & 62597 & 50199 & -20\% \\
\noalign{\smallskip}
\hline
\end{tabular} \\
\footnotesize
*Golden Gate Bridge NB link's closest PeMS sensor is located after an off ramp and hence the field count does not reflect the full bridge traffic count. 
\end{table}

\begin{figure}
    \centering
    \begin{subfigure}{.4\textwidth}
        \includegraphics[width=\textwidth]{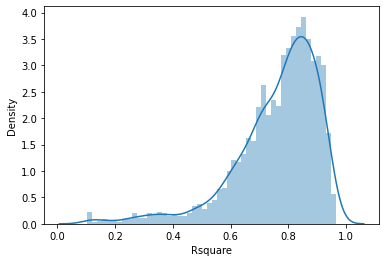}
        \label{fig:re}
        \caption{Distribution of R$^2$ values for the highway links. It follows a left skewed normal distribution with mean 0.75 and standard deviation 0.15}
    \end{subfigure}
    \begin{subfigure}{.55\textwidth}
        \includegraphics[width=\textwidth]{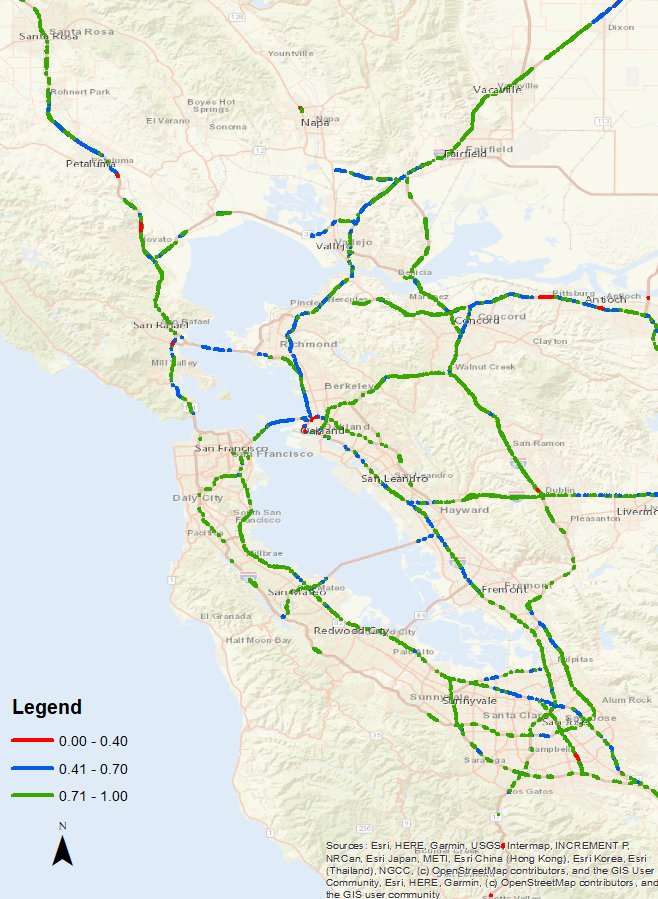}
        \label{fig:rsq}
        \caption{Spatial distribution of R$^2$ values of the links.}
    \end{subfigure}
    \caption{Evaluation metrics for highway link flows compared to PeMS sensors.}
    \label{fig:wholebaymetrics}
\end{figure}

Next, we evaluated R$^2$ for all links that have a corresponding PeMS  sensor in Bay Area. We were able to match 2061 links with mainline sensors and the resulting R$^2$ distribution is shown in Figure ~\ref{fig:wholebaymetrics}a. Of the total matched links, 72\% links have R$^2$ greater than 0.7 and 5\% have lower than 0.4 (Figure ~\ref{fig:wholebaymetrics}b).

% \begin{figure}[ht]
%     \includegraphics[width=.5\textwidth]
%     {sources/images/rsquare_b60.png}\hfill
%   \caption{Figure shows the R$^2$ flow values compared against PeMS sensors for highways in Bay Area.}
%   \label{fig:pems_flows}
% \end{figure}

\begin{figure}[ht]
  \includegraphics[width=.3\columnwidth]
    {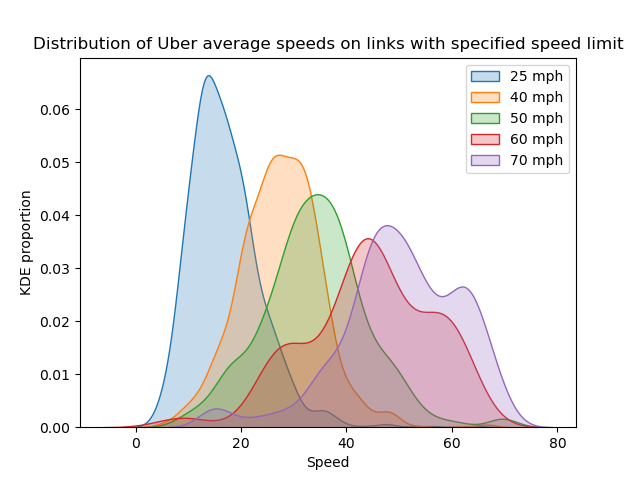}
  \includegraphics[width=.3\columnwidth]
    {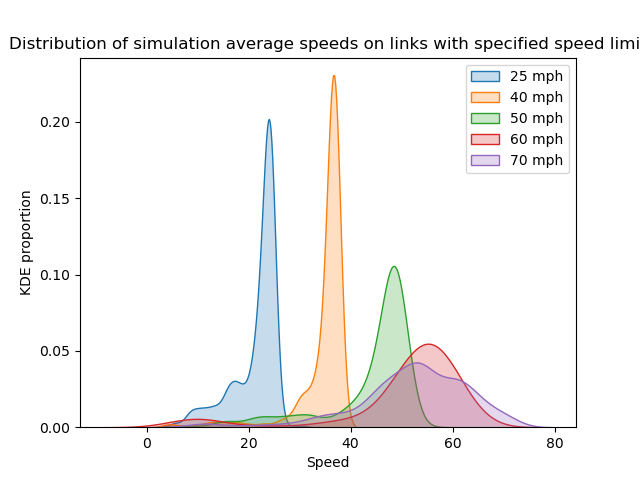}
  \caption{Average Uber (left) and Mobiliti (right) speeds across all speed limits 8am to 9am in the morning.}
  \label{fig:stage2a}
\end{figure}

\begin{figure}[ht]
  \includegraphics[width=.3\columnwidth]
    {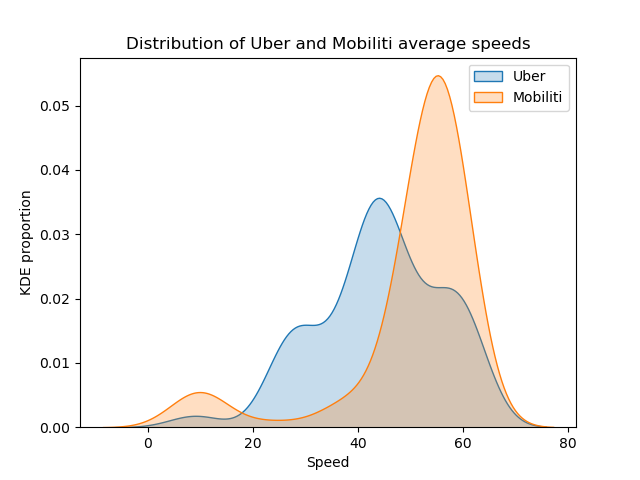}
  \includegraphics[width=.3\columnwidth]
    {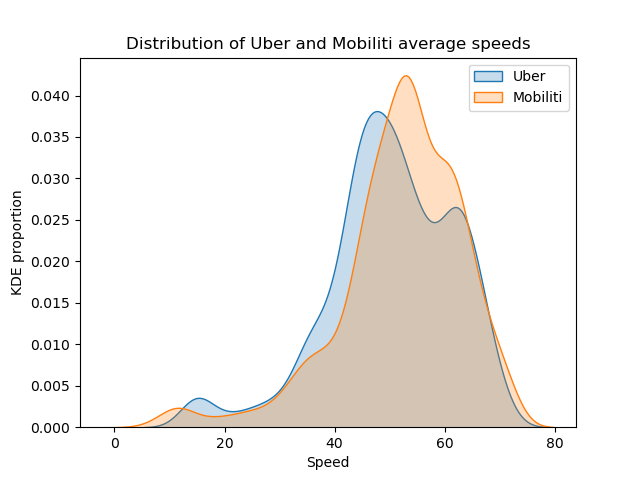}
  \caption{Kernel density comparing  Mobiliti  and Uber speed distributions at 60 mph (left) and 70 mph (right).}
  \label{fig:stage2b}
\end{figure}

Stage 2 in the validation procedure is speed comparison with a) Uber Movement data for San Francisco city streets, and b) PeMS speed data for Bay Area highways. For Uber Movement, speed data for San Francisco region for Q4, 2019 \cite{noauthor_uber_nodate} is used for comparison. Links from Uber's network were matched to Mobiliti links for a total of 139,495 links (20\% of total) in the simulation. The speeds were compared for 8 am to 9 am for different speed limits. Figure~\ref{fig:stage2a} shows the average speeds from Mobiliti and Uber across all speed limits and Figure~\ref{fig:stage2b} shows the speed distributions from Mobiliti and Uber on links with 60 mph and 70 mph speed limit.

\begin{figure}[ht]
  \includegraphics[width=.32\columnwidth]
    {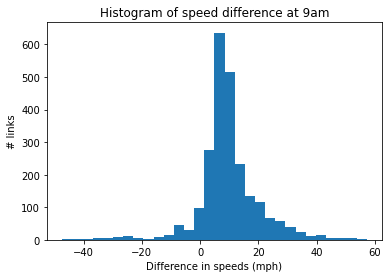}
  \includegraphics[width=.32\columnwidth]
    {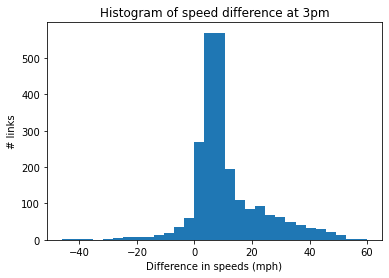}
  \caption{Figure shows the histogram of speed difference between simulation and PeMS at 9am and 3pm for highways in Bay Area.}
  \label{fig:stage2c}
\end{figure}

\begin{figure}[ht]
  \includegraphics[width=.32\columnwidth]
    {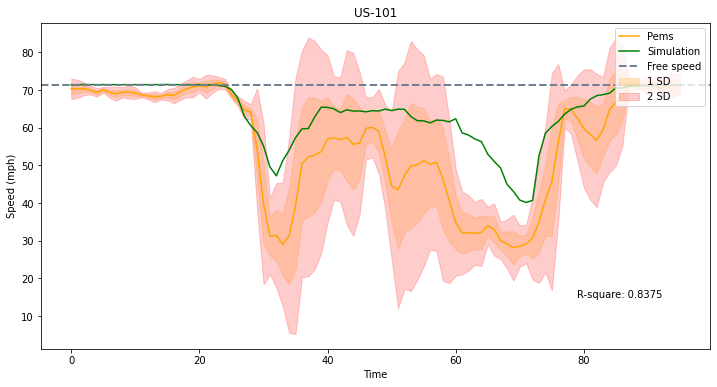}\hfill
  \includegraphics[width=.32\columnwidth]
    {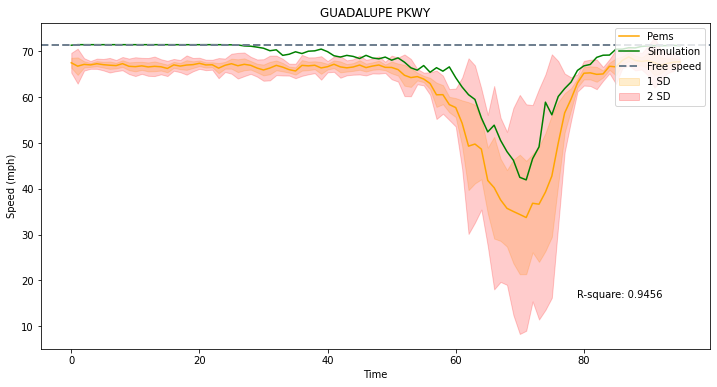}\hfill
  \includegraphics[width=.32\columnwidth]
    {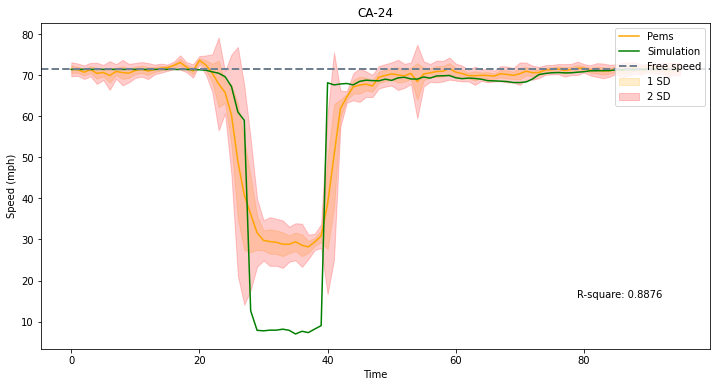}\hfill
  \includegraphics[width=.32\columnwidth]
    {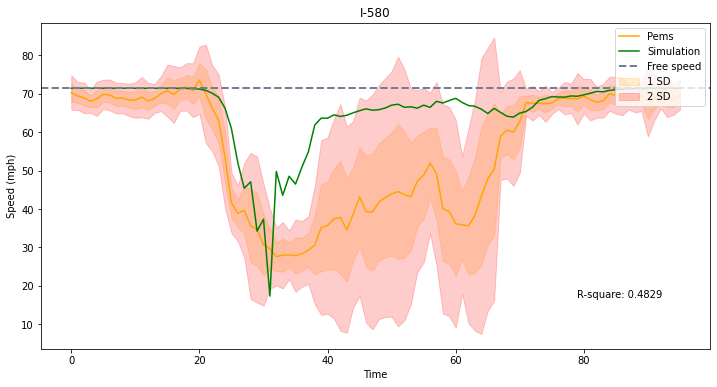}\hfill
  \includegraphics[width=.32\columnwidth]
    {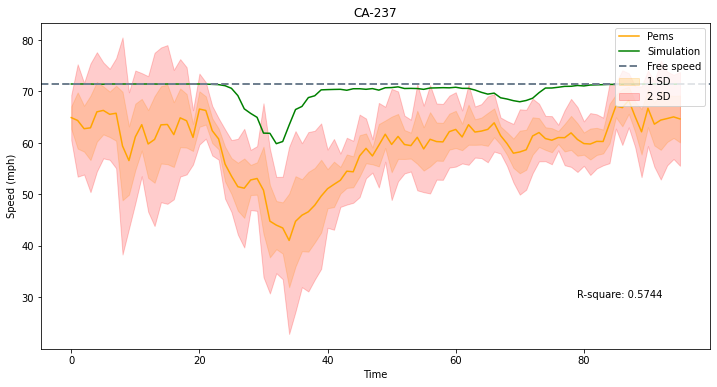}\hfill
  \includegraphics[width=.32\columnwidth]
    {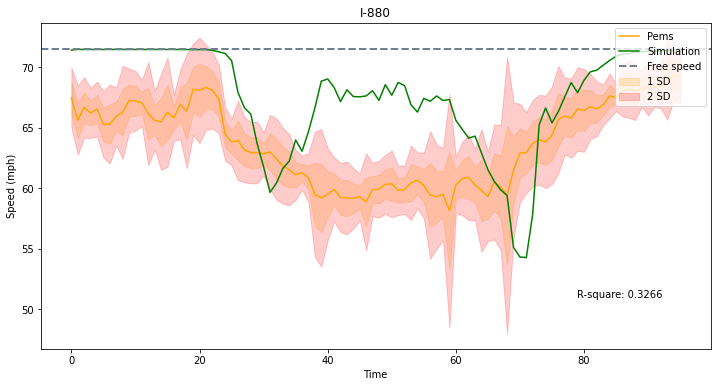}\hfill
  \caption{Simulation and PeMS speed profile for highways. Yellow line represents the average weekday values from PeMS. The first and second standard deviation bands are shown around the average. Green line represents the simulation speed for a typical day.}
  \label{fig:stage2d}
\end{figure}

Next we compared highway links with PeMS speed profiles for the 2061 matched links. Figure \ref{fig:stage2c} shows the difference in speeds between simulation and PeMS at 9 am and 3pm for a weekday. Most links are within $\pm20$ mph difference. Further, R$^2$ values were evaluated for all links to understand the time series trends. Figure \ref{fig:stage2d} shows a time series comparison for six highway links. Of the total, 55\% of highway links have R$^2$ greater than 0.4. We plan to improve the speed models to reflect time series trends closer to real world data in the future.

Stage 3 is system level comparisons, network validation, and error checking. Model visualization is used to check for unusual activities in traffic flows and odd roadway network attributes. Error checking and model verification consist of several smaller tasks such as checks for link geometry and connectivity, link speeds, and ramp and intersection geometry.  Since our travel demand data was obtained from SFCTA, which conducts their own validation, we did not conduct additional behavior checks.  We conducted system metric checks for VMT and total demand and validated them against the 2017 Environmental Impact report for the Bay Area \cite{noauthor_environmental_nodate} in Table~\ref{tab:stage3}.

\begin{table}[h]
\small
\centering
\caption{System level metrics}
\label{tab:stage3}      
\begin{tabular}{cccc}
\hline \textbf{Metric} & \textbf{Simulation} & \textbf{Field Data} & \textbf{Relative Error(\%)}\\
\hline
\noalign{\smallskip}
VMT & 146,546,360 & 158,406,800 & -7 \\
Daily Trips & 19,167,301 & 21,227,800 & -10 \\
\noalign{\smallskip}
\hline
\end{tabular}
\end{table}
\section{Conclusion}
Vehicles are rapidly gaining the ability to utilize up-to-date road congestion information to re-route their paths during their trips through smartphone navigation apps.  In this paper, we presented a computational methodology to model varying degrees of dynamic rerouting in a large-scale transportation system using the Mobiliti high-performance parallel discrete event simulator.  We described updates to our link actor model to capture the effects of link congestion, timing constraints, and storage capacity constraints.  We have detailed the implementation of new VehicleController actors and the events required to update their knowledge of the system state and service dynamic rerouting requests.  Because we have designed our simulator to scale over distributed memory parallel computing platforms, we can simulate one model day of the San Franscisco Bay Area with 19 million vehicle trips and 50 percent dynamic rerouting penetration over a road network with 0.5 million nodes and 1 million links in three minutes of simulation time.  We conducted an analysis of system-level impacts when varying the dynamic rerouting penetration rate at 10\% increments and examined the varying effects on different functional classes and geographical regions.  Finally, we presented a validation of the simulation results compared to real world data sources.  % There are many ways we could extend our simulator to capture more detailed vehicle behavior than the currently implemented mesoscopic link congestion model. Since the added complexity of more detailed modeling will likely reduce computational performance, one of our main objectives in the near future is to to determine the best way to balance model enhancements with computational performance.

\begin{acks}
This report and the work described were sponsored by the U.S. Department of Energy (DOE) Vehicle Technologies Office (VTO) under the Big Data Solutions for Mobility Program, an initiative of the Energy Efficient Mobility Systems (EEMS) Program. The following DOE Office of Energy Efficiency and Renewable Energy (EERE) managers played important roles in establishing the project concept, advancing implementation, and providing ongoing guidance: David Anderson and Prasad Gupta.  This research used resources of the National Energy Research Scientific Computing Center, a DOE Office of Science User Facility supported by the Office of Science of the U.S. Department of Energy under Contract No. DE-AC02-05CH11231.

\end{acks}

\bibliographystyle{ACM-Reference-Format}
\bibliography{references}

\end{document}